\documentclass[10pt,journal,compsoc]{IEEEtran}

\usepackage{bigstrut,array,multirow,tabularx}
\usepackage{diagbox}
\usepackage{slashbox}
\usepackage{makecell}
\usepackage{amsthm}
\usepackage{tikz}
\usepackage[super]{nth}
\usepackage{algorithm}
\usepackage{algorithmic}
\usepackage{graphicx}
\usepackage{verbatim}
\ifCLASSOPTIONcompsoc
    \usepackage[caption=false, font=normalsize, labelfont=sf, textfont=sf]{subfig}
\else
\usepackage[caption=false, font=footnotesize]{subfig}
\fi
\usetikzlibrary{patterns}
\usepackage{amsmath}
 
\usepackage{amssymb}
\usepackage{amsthm}
\usepackage[normalem]{ulem}
\usepackage[hidelinks]{hyperref}
\usepackage[nameinlink, capitalise, noabbrev]{cleveref}
\usepackage{mathtools}
\usepackage{dsfont}

\DeclarePairedDelimiter\ceil{\lceil}{\rceil}

\usepackage{tikz}
\usepackage{pgfplots} 
\pgfplotsset{compat=1.16} 
\usepackage{bigstrut,array,multirow,tabularx}
\usepackage{caption}
\usepackage{slashbox}

\newtheorem{definition}{Definition}[section]
\newtheorem{theorem}{Theorem}[section]

\newcounter{example}[section]
\newenvironment{example}[1][]{\refstepcounter{example}\par\medskip
   \noindent \textbf{Example~\theexample. #1} \rmfamily}{\medskip}

\usepackage{amsmath}

\usepackage{amsfonts}

\usepackage{multirow}
\usepackage{multicol}
\usepackage{caption}
\usepackage{array}
\usepackage{booktabs}
\usepackage{graphicx}

\theoremstyle{theorem}

\definecolor{bblue}{HTML}{FFE333}
\definecolor{rred}{HTML}{7FFF00}
\definecolor{ggreen}{HTML}{87CEEB}

\ifCLASSOPTIONcompsoc
  \usepackage[nocompress]{cite}
\else
  \usepackage{cite}
\fi
\ifCLASSINFOpdf
\else
\fi
\begin{document}
%
\title{Centrality-Based Node Feature Augmentation for Robust Network Alignment}


\author{Jin-Duk Park, Cong Tran, Won-Yong Shin, {\em Senior Member}, {\em IEEE}, and Xin Cao, {\em Member}, {\em IEEE}
\IEEEcompsocitemizethanks{\IEEEcompsocthanksitem J.-D. Park is with the School of Mathematics and Computing (Computational Science and Engineering), Yonsei University, Seoul 03722, Republic of Korea.
 \protect\\
E-mail: jindeok6@yonsei.ac.kr
\IEEEcompsocthanksitem C. Tran is with the Faculty of Information Technology, Posts and Telecommunications Institute of Technology, Hanoi 100000, Vietnam. \protect\\
E-mail: congtt@ptit.edu.vn

\IEEEcompsocthanksitem W.-Y. Shin is with the School of Mathematics and Computing (Computational Science and Engineering), Yonsei University, Seoul 03722, Republic of Korea, and the Graduate School of Artificial Intelligence, Pohang University of Science and Technology (POSTECH), Pohang 37673, Republic of Korea.\protect\\
E-mail: wy.shin@yonsei.ac.kr

\IEEEcompsocthanksitem X. Cao is with the School of Computer Science and Engineering, The University of New South Wales, Sydney 2052, Australia. \protect\\
E-mail: xin.cao@unsw.edu.au

(Corresponding author: Won-Yong Shin.)}}


%
%

\markboth{}%
{Shell \MakeLowercase{\textit{et al.}}: Bare Demo of IEEEtran.cls for Computer Society Journals}
%



%

\IEEEtitleabstractindextext{%
\begin{abstract}
Network alignment (NA) is the task of discovering node correspondences across multiple networks. Although NA methods have achieved remarkable success in a myriad of scenarios, their effectiveness is not without additional information such as prior anchor links and/or node features, which may not always be available due to privacy concerns or access restrictions. To tackle this challenge, we propose \textsf{Grad-Align+}, a novel NA method built upon a recent state-of-the-art NA method, the so-called Grad-Align, that {\em gradually} discovers a part of node pairs until all node pairs are found. In designing \textsf{Grad-Align+}, we account for how to {\em augment node features} in the sense of performing the NA task and how to design our NA method by maximally exploiting the augmented node features. To achieve this goal, \textsf{Grad-Align+} consists of three key components: 1) {\em centrality}-based node feature augmentation (CNFA), 2) {\em graph neural network (GNN)}-aided embedding similarity calculation alongside the augmented node features, and 3) gradual NA with similarity calculation using {\em aligned cross-network neighbor-pairs (ACNs)}. Through comprehensive experiments, we demonstrate that \textsf{Grad-Align+} exhibits (a) the superiority over benchmark NA methods, (b) empirical validations as well as our theoretical findings to see the effectiveness of CNFA, (c) the influence of each component, (d) the robustness to network noises, and (e) the computational efficiency.
\end{abstract}

}

\maketitle

\IEEEdisplaynontitleabstractindextext

%
\IEEEpeerreviewmaketitle

\section{Introduction}\label{sec:introduction}

\subsection{Background and Motivation}
\label{sec 1.1}

\IEEEPARstart{M}{u}ltiple social networks have become an integral part of our daily lives, serving various real-world applications \cite{zhang2016final, trung2020adaptive, man2016predict,chen2020cone}. However, a primary challenge on network analyses is that the user accounts across multiple networks are mostly isolated without any correspondence relationships \cite{yang2022anchor}, which hinders the development of cross-network applications. In this regard, the so-called {\it network alignment (NA)}, the task of discovering node correspondences across different networks, is often the very first step to perform downstream machine learning (ML) tasks on multiple networks, leading to more precise network analyses \cite{zhang2016final, park2023power, chen2020cone, heimann2018regal}. For example, by discovering the correspondence between Twitter and Foursquare networks of the same user, one can improve the performance of friend/location recommendations for Foursquare users whose social connections and activities can be very sparse \cite{kong2013inferring}. In addition to social network analyses, NA has been proven to be promising for solving ML problems across a wide range of fields such as computer vision, bioinformatics, web mining, and chemistry \cite{zhang2016final, trung2020adaptive}. As an example, in bioinformatics, aligning tissue-specific protein-protein interaction (PPI) networks can facilitate candidate gene prioritization \cite{ni2014inside}.

\IEEEpubidadjcol

Despite the remarkable success of NA methods \cite{chen2020cone, heimann2018regal, man2016predict,  zhou2018deeplink, trung2020adaptive, du2019joint, park2023power,zhang2016final, emmert2016fifty}, their effectiveness is not guaranteed without supervision data (i.e., prior anchor links) and/or node features. Nonetheless, such additional information may not always be available in real-world applications due to users' privacy and high-cost issues \cite{du2020cross, hsu2017unsupervised, ren2019meta}. For instance, cross-network anchor link labeling in social networks involves the arduous task of manually pairing user accounts across different networks and verifying user backgrounds, which can be time-consuming and labor-intensive \cite{ren2019meta}.

\begin{figure} 
    \centering
  \subfloat[\label{fig1a}]{%
       \includegraphics[width=0.47\columnwidth]{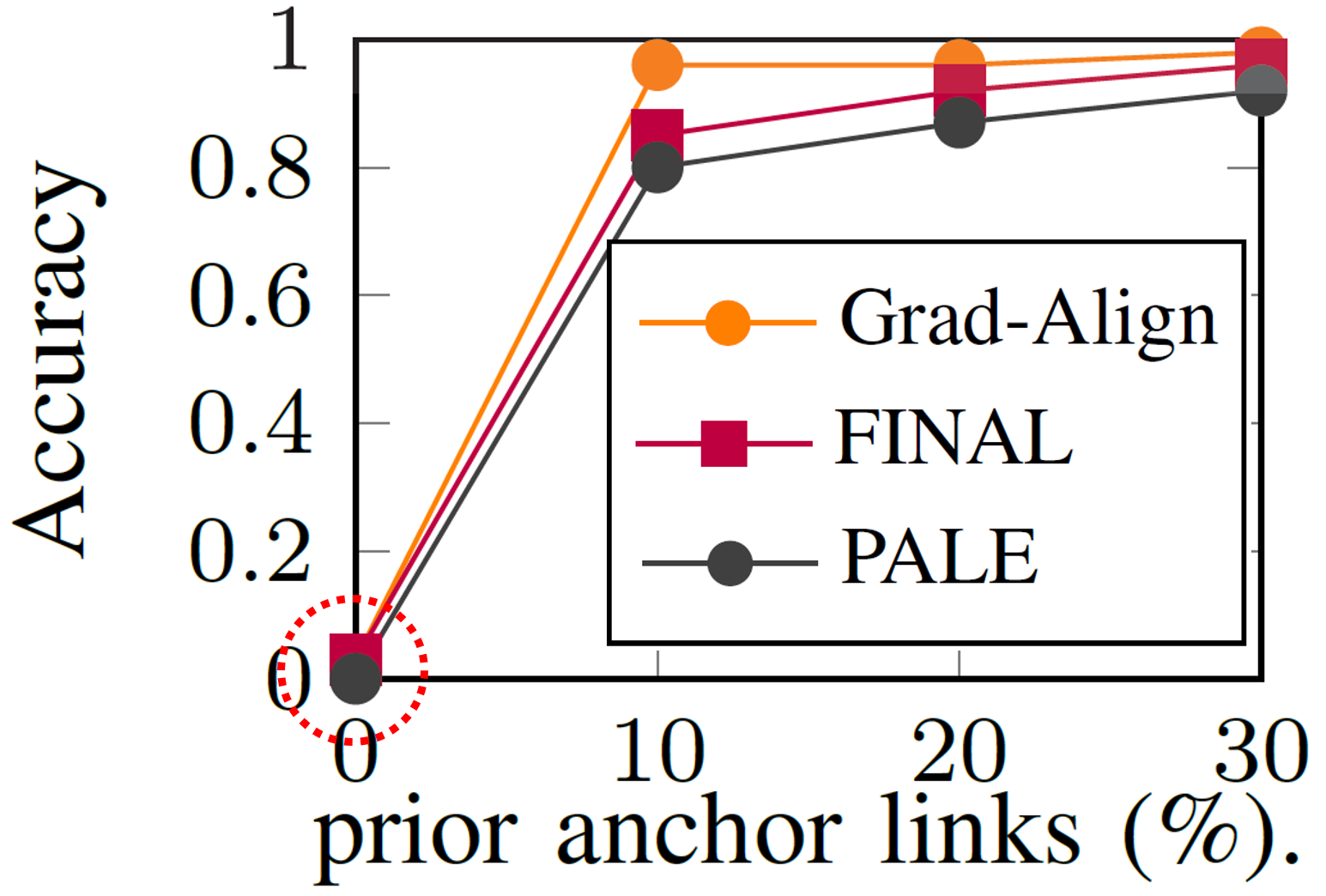}}
    \hfill
  \subfloat[\label{fig1b}]{%
        \includegraphics[width=0.46\columnwidth]{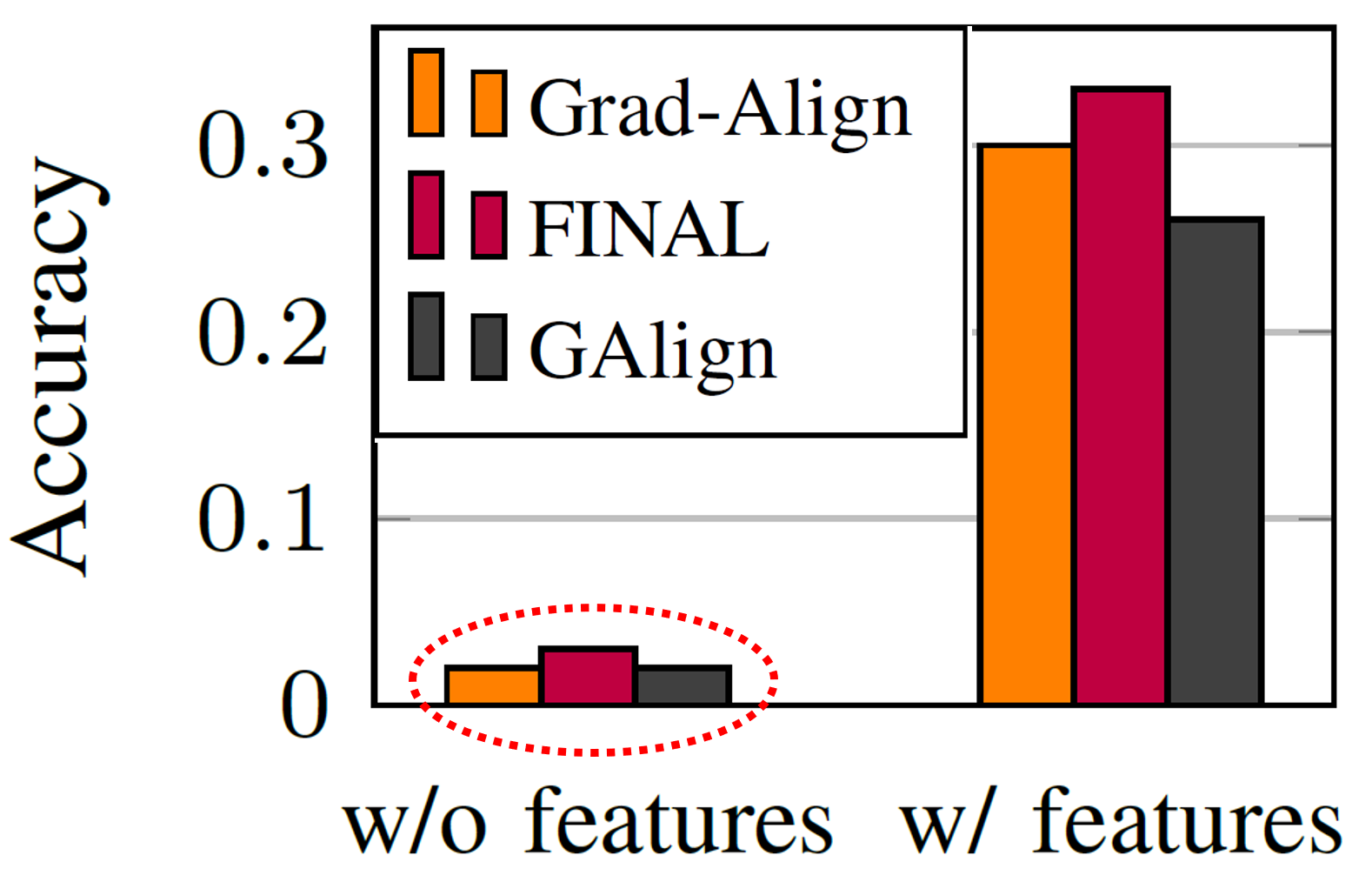}}
    \hfill
  \caption{Alignment accuracy (a) for the scenario where a portion of prior anchor links vary on Facebook vs. Twitter and (b) for the scenario with and without node features on Douban Online vs. Douban Offline.}
  \label{fig1} 
\end{figure}

Our study is motivated by the observation that the state-of-the-art performance of existing NA methods is significantly degraded when prior anchor links and/or original node features (also known as node attributes) are unavailable. Fig. \ref{fig1a} illustrates the alignment accuracy for the scenario where a portion of prior anchor links vary on Facebook vs. Twitter. As depicted in the figure, NA methods designed by leveraging prior anchor link information (e.g., PALE \cite{man2016predict}, FINAL \cite{zhang2016final}, and Grad-Align \cite{trung2020adaptive}) tend to reveal high alignment accuracies when prior anchor links are used (albeit slightly). However, surprisingly, when anchor links are no longer available, all of these NA methods perform very poorly while showing accuracies much lower than 0.1 (see the red circle depicted in Fig. \ref{fig1a}). Moreover, Fig. \ref{fig1b} illustrates the alignment accuracy for the scenario with and without node features on Douban Online vs. Douban Offline. As visualized in the figure, NA methods making use of node features (e.g., GAlign \cite{trung2020adaptive}, FINAL \cite{zhang2016final}, and Grad-Align \cite{park2023power}) also reveal drastically degraded performance for non-attributed network settings such as the Douban dataset with the removal of node features (see the red circle depicted in Fig. \ref{fig1b}). 

\subsection{Main Contributions}

Based on the aforementioned practical challenges posed by existing NA methods, a natural question arising is ``How to avoid such dependencies on additional information including prior anchor links and node features in the NA task?" To answer this question, we aim at developing an NA method that is capable of achieving outstanding performance even in the absence of such additional information, by taking advantage of \textit{node feature augmentation (NFA)} that suits the NA problem. To this end, we first outline two design challenges that must be addressed when building a new NA method.
\begin{itemize}
    \item \textbf{NFA:} how to effectively augment node features in the sense of performing the NA task;
    \item \textbf{Exploitation of the augmented node features in NA:} how to design our NA method by maximally exploiting the augmented node features. 
\end{itemize}

\begin{figure}
    \centering
    \includegraphics[scale=0.5]{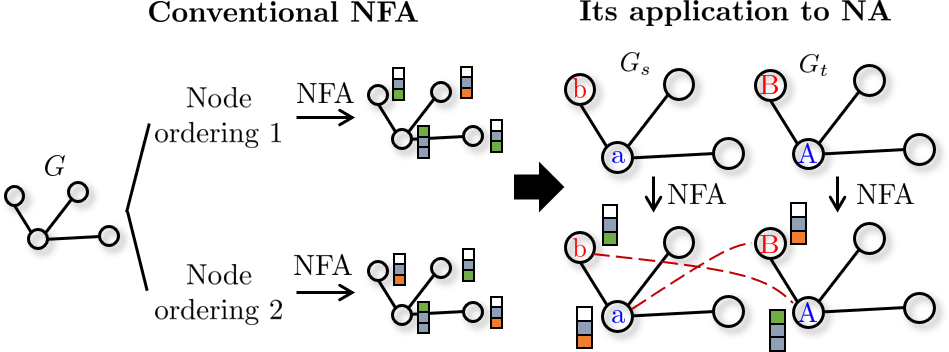}
    \caption{An example illustrating a na\"ive NFA approach for NA. Here, (a,A) and (b,B) are ground truth cross-network node pairs, and the red dashed curves denote the (incorrectly) discovered node correspondences via NA.}
    \label{fig:intro_exam}
\end{figure}

\textbf{(\underline{Idea 1})} 
When the node features are not originally available, we consider a strategy of employing NFA techniques so that NA methods perform robustly even in the absence of additional information. We start by highlighting that NFA for NA poses distinct technical difficulties that must be approached differently from the implementation of NFA in other ML tasks on graphs. In our study, we present the following three crucial guiding principles of NFA that enable us to achieve satisfactory performance on NA: 1) \textit{expressiveness}, 2) \textit{permutation invariance}, and 3) \textit{consistency}. More specifically, augmented node features must be 1) sufficiently expressive to distinguish one node from others in a given network, 2) independent of varying node orderings, and 3) generated in such a way that the same node correspondence across different networks exhibits similar node features. However, simply adopting existing NFA approaches designed for general graph learning tasks \cite{kong2020flag, zhu2021graph, wang2020nodeaug} may not suffice the aforementioned guiding principles. Existing NFA methods rely primarily on perturbation-based techniques along with {\it originally given} node features \cite{kong2020flag, zhu2021graph, wang2020nodeaug} and do not account for scenarios where node features are totally unavailable. Moreover, NFA approaches utilizing embedding via topological regularization \cite{song2021topological} and spectral pre-coloring \cite{feldman2022weisfeiler} do not meet the permutation invariance principle, which can lead to largely different node features in a ground truth cross-network node pair. Fig. \ref{fig:intro_exam} illustrates an instance of source and target networks with arbitrary node orderings, denoted by $G_s$ and $G_t$, where node features are augmented based on classical network embedding techniques such as DeepWalk \cite{perozzi2014deepwalk}, LINE \cite{tang2015line}, and node2vec \cite{grover2016node2vec}, as in \cite{song2021topological}. As depicted in the figure, the augmented node features for each ground truth node pair (i.e., (a,A) or (b,B)) across isomorphic networks are not identical, despite having the same topological structure. This violates the consistency principle in NA, indicating that a ground truth node pair shall share consistent node features \cite{zhang2016final,trung2020adaptive,park2023power}, which thus yields the undesirable discovery of node pairs. Furthermore, while some existing methods \cite{zhang2016final,tang2023identifying} have partially utilized NFA for NA by leveraging only the degree of nodes, such limited approaches have often resulted in suboptimal performance \cite{zhang2016final}. Thus, despite the above-described contributions, the potential of NFA for NA has not yet been thoroughly explored, particularly in terms of three crucial dimensions: expressiveness, permutation invariance, and consistency. To fundamentally address this problem, we devise \textit{centrality-based NFA (CNFA)}, a new NFA technique suitable for the NA task that satisfies all of the three guiding principles. CNFA is composed of 1) centrality calculation, 2) centrality selection based on our own {\it centrality selection score} computed prior to performing NFA, and 3) equal-width binning. Our CNFA technique is capable of judiciously encoding each node's {\it centrality} by preserving the expressive and permutation invariance properties.

\begin{figure*}
    \centering
    \includegraphics[width=0.9\textwidth]{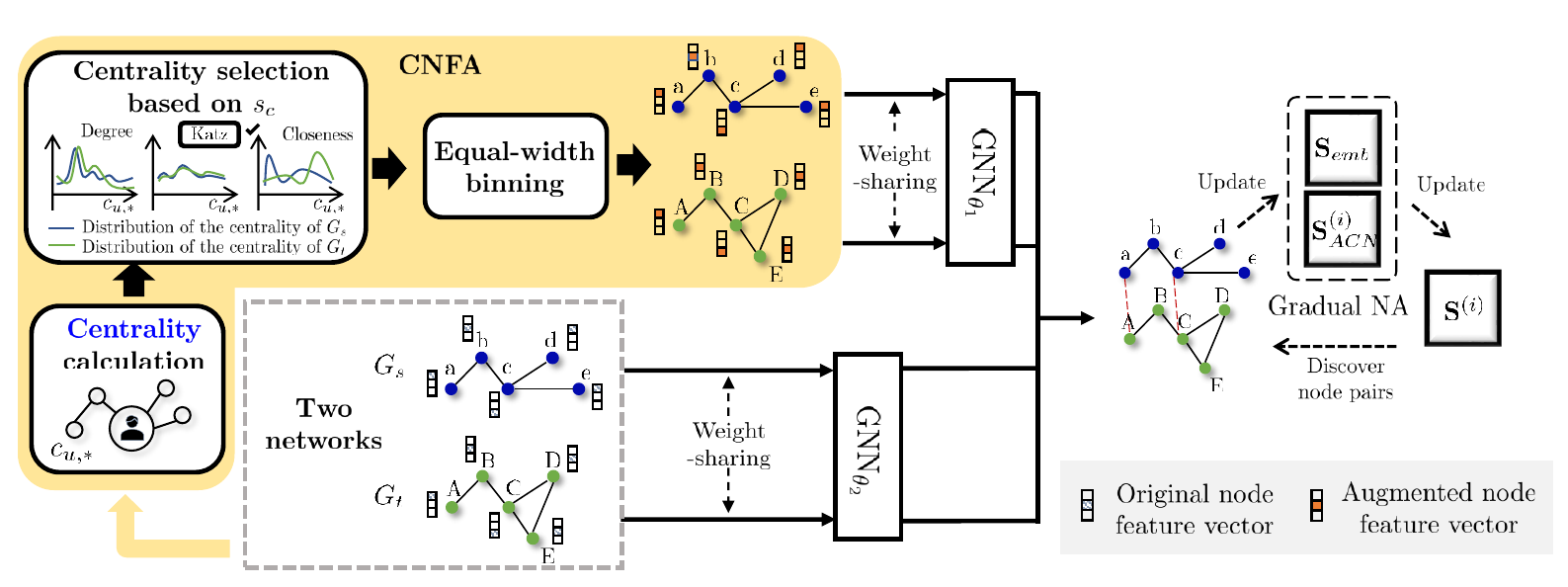}
    \caption{The schematic overview of our \textsf{Grad-Align+} method.}
    \label{fig:overview}
    \vspace{-3mm}
\end{figure*}

\textbf{(\underline{Idea 2})} 
We propose \textsf{Grad-Align+}, a new NA method that makes full use of the augmented node features based on our CNFA technique and is no longer vulnerable to the absence of additional information. \textsf{Grad-Align+} is built upon a recent state-of-the-art NA method, named as Grad-Align \cite{park2023power}, that {\it gradually} discovers node pairs by harnessing information enriched through \textit{interim} discovery of node correspondences during the node matching.\footnote{Note that CNFA most benefits gradual NA methods (e.g., Grad-Align \cite{park2023power}) by helping find correct node pairs, especially in the early stage of gradual node matching, which will be empirically validated in Section \ref{rq4}.} In \textsf{Grad-Align+}, we present our new embedding similarity calculation using graph neural networks (GNNs) as a powerful means to extract useful low-dimensional features. Although GNNs can learn expressive node representations even without original node features by randomly assigning a one-hot encoded vector to each node in general graph learning tasks \cite{DBLP:conf/iclr/KipfW17}, this is not appropriate as long as the NA task is concerned due to the violation of the consistency principle. Thus, instead of using one-hot encoded vectors, we generate node representations through GNNs with the aid of the augmented node features via CNFA. Finally, we characterize a new measure, the so-called {\it ACN similarity}, that represents similarities between cross-network nodes using the information of {\it aligned cross-network neighbor-pairs (ACNs)} \cite{park2023power}. Then, we gradually discover node pairs by calculating both the multi-layer embedding similarity upon the node representations and the ACN similarity.

To validate the effectiveness of \textsf{Grad-Align+}, we conduct theoretical analyses as well as empirical evaluations. Through comprehensive experiments on three real-world and three synthetic benchmark datasets, including one large-scale dataset consisting of more than 34k nodes, we demonstrate that the proposed \textsf{Grad-Align+} method significantly and consistently outperforms the best NA competitor up to the gain of 69.80\% in alignment accuracy. Additionally, we theoretically analyze 1) how the expressiveness of node representations via GNNs can be enhanced using CNFA and 2) how the augmented features can benefit the resulting vector representations via GNNs. We also analyze the computational complexity of \textsf{Grad-Align+}, which shows a \textit{linear scaling }in the number of edges.

The main contributions of this paper are summarized as follows:

\begin{itemize}

\item \textbf{Important observations:} We point out the limitations of existing NA methods on achieving satisfactory performance in the absence of prior anchor links and/or node features. To the best of our knowledge, there is no prior attempt to tackle these practical challenges in the literature.

\item \textbf{Novel methodology:} We propose \textsf{Grad-Align+}, a novel NA method that is robust to the unavailability of additional information. Precisely, our \textsf{Grad-Align+} method distinguishes itself from Grad-Align \cite{park2023power} in three perspectives: 1) CNFA that encodes the centrality of each node, 2) GNN-based embedding similarity calculation along with the \textit{augmented} node features, and 3) gradual NA with \textit{ACN similarity}, a new measure that represents the similarity between cross-network nodes.

\item \textbf{Analysis and evaluation:} We demonstrate (a) the superiority of \textsf{Grad-Align+} over seven state-of-the-art NA methods by a significant margin, (b) the impact of CNFA and key parameters in \textsf{Grad-Align+}, (c) the influence of each component in \textsf{Grad-Align+}, and (d) the robustness of \textsf{Grad-Align+} to network noise. We also provide rigorous theoretical foundations for validating the effectiveness of core components in \textsf{Grad-Align+}.

\end{itemize}
We refer to Appendix C for a comprehensive review of the related work on NFA and NA methods.

\section{Methodology}
\label{section 3}
In this section, as a basis for the proposed \textsf{Grad-Align+} method in Section \ref{section 4}, we first present our network model with basic assumptions and define the NA problem. Then, we describe an overview of our \textsf{Grad-Align+} method using CNFA as a core component. The schematic overview of our proposed method is illustrated in Fig. \ref{fig:overview}. We refer to Appendix A for an overview of the notations used throughout the paper.

\subsection{Network Model and Basic Assumptions}
\label{sub3.1}

We consider source and target networks to be aligned, $G_s$ and $G_t$, respectively. For simplicity, we assume that both $G_s$ and $G_t$ are undirected and unweighted. For the source network $G_s = (\mathcal{V}_s,\mathcal{E}_s,\mathcal{X}_s)$, $\mathcal{V}_s$ is the set of nodes (or equivalently, vertices) in $G_s$ whose cardinality is $n_s$; $\mathcal{E}_s$ is the set of edges between nodes in $\mathcal{V}_s$; and $\mathcal{X}_s \in \mathbb{R}^{n_s \times d_o}$ is the set of node feature (or equivalently, node attribute) vectors, where $d_o$ is the dimensionality of the vector space. For the target network $G_t = (\mathcal{V}_t,\mathcal{E}_t,\mathcal{X}_t)$, notations similarly follow those in $G_s$ with $|\mathcal{V}_t|=n_t$ and $\mathcal{X}_t \in \mathbb{R}^{n_t \times d_o}$.\footnote{As in \cite{trung2020adaptive,zhang2016final,du2019joint,park2023power}, the dimension of each node feature vector, $d_o$, for both networks is assumed to be the same.}  In our study, we further denote prior seed nodes in source and target networks as $\tilde{\mathcal{V}}_s^{(0)}$ and $\tilde{\mathcal{V}}_t^{(0)}$, respectively, where edges connecting cross-network prior seed nodes correspond to the ground truth prior anchor links for NA \cite{park2023power}. Here, node features $\mathcal{X}_s$ and $\mathcal{X}_t$ and prior seed nodes $\tilde{\mathcal{V}}_s^{(0)}$ and $\tilde{\mathcal{V}}_t^{(0)}$ are \textit{optional} depending on real-world network circumstances.

\subsection{Problem Definition}

    In this subsection, we define the problem of NA for given two networks $G_s$ and $G_t$ formally as follows.

    \begin{definition}[NA] \cite{park2023power, du2019joint}: Given source and target networks
    $G_s=(\mathcal{V}_s,\mathcal{E}_s,\mathcal{X}_s)$ and $G_t=(\mathcal{V}_t,\mathcal{E}_t,\mathcal{X}_t)$, NA aims at discovering one-to-one node mapping $\pi:\mathcal{V}_s \rightarrow \mathcal{V}_t$, where $\pi(u) = v$ and $\pi^{-1}(v) = u$ for $u \in \mathcal{V}_s$ and $v \in \mathcal{V}_t$.
    \end{definition}


 
\subsection{Overview of \textsf{Grad-Align+}}

In this subsection, we explain our methodology alongside the overview of our \textsf{Grad-Align+} method. \textsf{Grad-Align+} is basically built upon Grad-Align \cite{park2023power}, a recent state-of-the-art NA method, which {\it gradually} discovers node pairs by harnessing information enriched through the interim discovery of node correspondences (referred to as ACNs) to more accurately find the next correspondences during the node matching. Upon the gradual NA framework, \textsf{Grad-Align+} is designed in the sense of taking advantage of {\it NFA} so that it is quite robust to the settings where prior anchor links and/or node features are unavailable. Our \textsf{Grad-Align+} is basically composed of three main phases: 1) CNFA, 2) GNN-based embedding similarity calculation, and 3) gradual NA with ACN similarity calculation. 


{\bf (Phase 1: CNFA)} In our study, we present {\it CNFA} for the NA task. This idea is motivated from the fact that other NFA techniques such as embedding via topological regularization \cite{song2021topological} and spectral pre-coloring \cite{feldman2022weisfeiler} are not permutation-invariant. Such existing NFA methods can allow the augmented node features to vary based on the node ordering of networks. That is, without knowledge of the node correspondence, two nodes in a ground truth cross-network node pair can exhibit largely different node features based on the arbitrary node ordering in each network. This poses a technical challenge for NA since the consistency assumption across the networks is violated, ultimately leading to significant performance degradation. Thus, it is of paramount importance to augment node features according to a new design principle that should maintain \textit{permutation invariance} in order to guarantee satisfactory performance of NA.

To this end, we propose the CNFA that ensures three key guiding principles for each node's {\it centrality}: 1) expressiveness, 2) permutation invariance, and 3) consistency (see Section \ref{section 4.1.1} for further details). In CNFA, we start by calculating widely used centrality measures, including degree, eigenvector \cite{bonacich1972factoring}, Katz \cite{katz1953new}, betweenness \cite{freeman1977set}, PageRank \cite{page1999pagerank}, and closeness \cite{freeman1978centrality}, for a given network. Next, as depicted in Fig. \ref{fig:overview}, we turn to selecting one of the centrality measures, which is crucial for achieving accurate NA results, as different centrality measures may reveal different levels of expressiveness for a given network topology \cite{rodrigues2019network, borgatti2005centrality}. For instance, we empirically examined that PageRank and closeness centralities are less expressive than others in the Zachary's Karate club network \cite{zachary1977information} (see Appendix B for a visualization example). In this context, rather than na\"ively choosing one of the centrality measures as in our conference version \cite{park2022gradalign+}, we newly devise a centrality selection score $s_c$, which is computed \textit{prior to} performing NFA without knowing the ground truth prior anchor links to find the best centrality measure for a given dataset. Specifically, we calculate $s_c$ based only on the distributions of each centrality measure of $G_s$ and $G_t$, then choose the one that has the highest $s_c$ as the best centrality measure. This centrality selection step plays a pivotal role in performing NA in reality without the ground truth node pairs, while guaranteeing robust performance regardless of experimental settings.
A detailed description of the score $s_c$ will be shown in Section \ref{section 4.1.1}. As the final step of CNFA, based on the selected centrality measure, we augment node features, where the centrality measure is discretized and encoded with a fixed dimension by employing the equal-width binning technique \cite{catlett1991changing}. Then, when the subscript $*$ represents $s$ and $t$ for source and target networks, respectively, it follows that
\begin{equation}
\begin{aligned}
  \hat{\bf x}_{u,*} = f_w(c_{u,*}),
\end{aligned}
\end{equation}
where $\hat{\bf x}_{u,*}$ is the $d$-dimensional augmented node feature vector of node $u$ in $G_*$, corresponding to the $u$-th element of augmented node feature matrix $\hat{\mathcal{X}}_* \in\mathbb{R}^{n_*\times d}$; $c_{u,*}$ is the centrality of node $u$ in $G_*$; and $f_w(.)$ is the equal-width binning function parameterized by binning width $w$. The detailed description on our CNFA phase will be shown in Section \ref{section 4.1.1}.

{\bf (Phase 2: GNN-based embedding similarity calculation)} We turn to calculating the multi-layer embedding similarity matrix via $L$-layer GNNs along with the augmented node feature matrix $\hat{\mathcal{X}}_*$ in Phase 1. First, if the two networks have \textit{original} node features $\mathcal{X}_*$, then we feed $\mathcal{X}_*$ into a GNN to calculate the hidden representation matrix at each layer for the network $G_*$:
 \begin{equation}
    \label{hid_rep_withorigin}
    \mathbf{H}_*^{(l)}=GNN_{\theta_1}(\mathbf{A}_*,\mathcal{X}_*),
\end{equation}
 where $GNN_{\theta_1}$ is the GNN model parameterized with $\theta_1$; ${\bf A}_*$ is the adjacency matrix (i.e., structural information) of $G_*$; and $\mathbf{H}_*^{(l)}$ is the hidden representation matrix at the $l$-th layer of $GNN_{\theta_1}$. Next, we feed the augmented node features $\hat{\mathcal{X}}_*$ into an \textit{additional} GNN model, $GNN_{\theta_2}$, to handle the heterogeneity of the feature vectors:
\begin{equation}
    \label{hid_rep}
    \hat{\mathbf{H}}_*^{(l)}=GNN_{\theta_2}(\mathbf{A}_*,\hat{\mathcal{X}}_*),
\end{equation}
 where $GNN_{\theta_2}$ is the GNN model parameterized with $\theta_2$; and $\hat{\mathbf{H}}_*^{(l)}$ is the hidden representation matrix at the $l$-th layer of $GNN_{\theta_2}$. Here, the model parameters $\theta_1$ and $\theta_2$ are trained by a \textit{layer-wise reconstruction loss} proposed in \cite{park2023power} to make each node representation more distinguishable. As illustrated in Fig. \ref{fig:overview}, using the $h$-dimensional hidden representations ${\bf H}_*^{(l)}\in\mathbb{R}^{n_*\times h}$ and $\hat{\bf H}_*^{(l)}\in\mathbb{R}^{n_*\times h}$ at each layer extracted from $GNN_{\theta_1}$ and $GNN_{\theta_2}$, respectively, we are capable of computing the multi-layer embedding similarity matrix as follows:
\begin{equation}
\label{emb_sim}
{\mathbf{S}}_{emb} = \sum_l \mathbf{H}_s^{(l)}{\mathbf{H}_t^{(l)~\top}} + \lambda \sum_l  \hat{\mathbf{H}}_s^{(l)}\hat{\mathbf{H}}_t^{(l)~\top},
\end{equation}
where $\lambda$ is a hyperparameter balancing between two terms in (\ref{emb_sim}). When the networks do not have original node features $\mathcal{X}_*$, we only use the second term in (\ref{emb_sim}) to calculate the multi-layer embedding similarity matrix.

{\bf (Phase 3: Gradual NA with ACN similarity calculation)} We describe how to discover node pairs gradually using a similarity matrix in each gradual step. In this phase, we repeatedly update the similarity matrix based on the updated ACNs \cite{park2023power} in each gradual step. In our study, rather than adopting prior approaches based on the Jaccard index \cite{du2019joint} and the Tversky similarity \cite{park2023power}, we devise a new measure, the so-called \textit{ACN similarity}, that represents similarities between cross-network nodes using the information of ACNs. Specifically, we formulate the ACN similarity as:
\begin{equation}
\label{ACN_sim}
    {\mathbf S}_{ACN}^{(i)}(u,v) = ACN_{u,v}^{(i)~p}, 
\end{equation}
where $ACN_{u,v}^{(i)~p}$ is the $p$-th power of the number of ACNs between node pair $(u,v)$ for $u\in\mathcal{V}_s$ and $v\in\mathcal{V}_t$ at the $i$-th iteration; and ${\bf S}_{ACN}^{(i)}(u,v)$ is the $(u,v)$-th element of the ACN similarity matrix ${\bf S}_{ACN}^{(i)}$. Finally, following the dual-perception similarity in Grad-Align \cite{park2023power}, we calculate the similarity matrix ${\bf S}^{(i)}$ as follows:
\begin{equation}
\label{overall_sim}
    {\mathbf S}^{(i)} = {\mathbf S}_{emb} \odot {\mathbf S}_{ACN}^{(i)},
\end{equation}
where $\odot$ indicates the element-wise matrix multiplication operator. The rest of gradual NA essentially follows that of \cite{park2023power}.

\section{Proposed Method: \textsf{Grad-Align+}}\label{section 4}
In this section, we elaborate on our \textsf{Grad-Align+} method along with theoretical analyses that justify our methodology. Additionally, we analyze the computational complexity of \textsf{Grad-Align+}.
\label{sec 4.1}


\subsection{Methodological Details of \textsf{Grad-Align+}}
We describe the three key phases in \textsf{Grad-Align+}. We refer to Appendix D for the pseudocode of the proposed \textsf{Grad-Align+} method.

\subsubsection{CNFA}
\label{section 4.1.1}
We first introduce CNFA, a technique for augmenting node features in the NA task. It is worth noting that CNFA is built upon three crucial guiding principles:
\begin{itemize}
    \item \textit{Expressiveness}: As a core principle of NA to identify node correspondences across different networks, the augmented node features should be sufficiently expressive to distinguish from other nodes in each network;
    \item \textit{Permutation invariance}: The augmented node features should be independent of varying node orderings, which is vital in the context of NFA for NA. To clarify, let $f_{NFA}$ denote an NFA function that augments a set of node features given a network and its node ordering as input. Then, for the network $\mathcal{G}$ with node orderings $\pi_1$ and $\pi_2$, it follows that $f_{NFA}(\mathcal{G}, \pi_1) =f_{NFA}(\mathcal{G}, \pi_2)$;
    \item \textit{Consistency}: The same node correspondence $(u,v)$ across two networks is assumed to have consistent node features, i.e., it should suffice to have $\hat{\bf x}_{u,s} \simeq \hat{\bf x}_{v,t}$.
\end{itemize}
\begin{figure} 
  \subfloat[\label{fig5a}]{%
       \includegraphics[width=0.47\columnwidth]{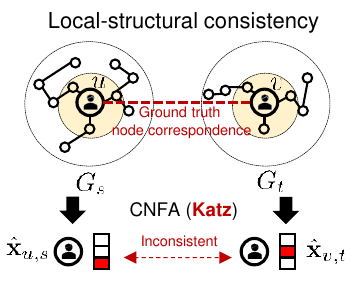}}
    \hfill
  \subfloat[\label{fig5b}]{%
        \includegraphics[width=0.46\columnwidth]{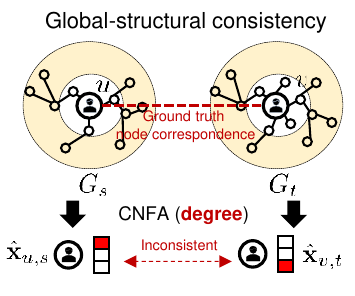}}
    \hfill
  \caption{Examples illustrating inconsistent node features based on (a) Katz centrality and (b) degree centrality. The structurally consistent area is highlighted in yellow.}

  \label{fig1} 
\end{figure}
The motivation behind our CNFA lies in the fact that real-world networks are often composed of nodes representing distinguishable structural connectivity \cite{barabasi1999emergence}. To capture this expressive property for NFA, we use widely-studied centrality measures, which indeed ensure that the augmented node features are effective in distinguishing between nodes in a network. In our study, various centrality measures are adopted to quantify the importance of nodes in a network; the centrality of node $i$ in $G_*$, denoted by $c_{i,*}$, can be expressed as:
\begin{itemize}
    \item Degree centrality: $c_{i,*} = |\mathcal{N}_i|$, where $|\mathcal{N}_i|$ is the cardinality of the set of neighbors of node $i$;
    \item Eigenvector centrality \cite{bonacich1972factoring}: $c_{i,*} =  \frac{1}{\alpha}\sum_{j \in \mathcal{N}_i}\mathbf{x}_{j,*}$, where $\alpha$ is a hyperparameter;
    \item Katz centrality \cite{katz1953new}: $c_{i,*} = \sum_{k = 1}^{\infty}\sum_{j \in \mathcal{N}_i}\alpha^k\mathbf{A}^k_{ij,*}+\beta$, where $\alpha$ and $\beta$ are a hyperparameters and ${\bf A}_{ij,*}$ is the $(i,j)$-th element of ${\bf A}_*$;
    \item Betweenness centrality \cite{freeman1977set}: $c_{i,*} = \sum_{{u,v} \in \mathcal{V}_*}\frac{\sigma(u,v|i)}{\sigma(u,v)}$, where $\sigma(u,v)$ is the number of the shortest $(u,v)$-paths and $\sigma(u,v|i)$ is the number of those paths passing through node $i$ other than $u$ and $v$;
    \item PageRank centrality \cite{page1999pagerank}: $c_{i,*} = \alpha\sum_{j \in \mathcal{V}_*}c_{j,*}+\frac{1-\alpha}{|\mathcal{V}_*|}$, where $\alpha \in [0,1)$ is a hyperparameter;
    \item Closeness centrality \cite{freeman1978centrality}: $c_{i,*} = \frac{|\mathcal{V}_*|-1}{\sum_{j=1}^{|\mathcal{V}_*|-1}d(i,j)}$, where $d(i,j)$ is the shortest path distance between nodes $i$ and $j$.
\end{itemize}
Note that all of these measures are {\it permutation-invariant}, and thus do not depend on the identity of each node, but rather on its position in a given network \cite{bloch2016centrality}.

Meanwhile, the same node correspondence across different networks often exhibits different structural patterns depending on datasets \cite{du2019joint, park2023power}. For instance, Figs. \ref{fig5a} and \ref{fig5b} illustrate the cross-network node pairs that have local-structural consistency and global-structural consistency, respectively. In this context, in the case of Fig. \ref{fig5a}, using centrality measures that capture global structural patterns (e.g., Katz centrality) will not be proper for augmenting consistent node features across the different networks; likewise, in the case of Fig \ref{fig5b}, using degree centrality for NFA is undesirable. We demonstrated that such an inappropriate choice of centrality can lead to severe performance degradation, especially when the original node feature is unavailable (see Table \ref{Q1table}). Therefore, selecting the most appropriate centrality measure is crucial from a practical standpoint. To attain this, we would like to present how to select the best centrality measure in terms of expressiveness and consistency for NFA so that CNFA is applicable in practice without the ground truth node pairs beforehand. To achieve this goal, we propose the centrality selection score $s_c$, which is computed prior to performing NFA. Here, $s_c$ is designed in such a way of having a high value if the two distributions of the centrality of $G_s$ and $G_t$ are sufficiently expressive and consistent to perform NFA for NA. To this end, we formalize $s_c$ as follows: 

\begin{equation}
\label{centrality selection score}
s_c = e^{\sigma^2_s+\sigma^2_t - \gamma D_{KL}(p_s||p_t)-1},
\end{equation}
where $p_*$ is the distribution of centrality over all nodes in $G_*$; $\sigma_*$ is the standard deviation of $p_*$; $D_{KL}(\cdot \| \cdot)$ is the Kullback--Leibler (KL) divergence of two distributions; and $\gamma$ is the hyperparameter balancing between the importance of expressiveness and consistency for NFA in NA. Here, smaller $\gamma$ places more emphasis on expressiveness, while larger $\gamma$ means more focus on consistency. We again note that, without this centrality selection using \eqref{centrality selection score}, the proposed \textsf{Grad-Align+} method does not guarantee robust performance in NA for all datasets.

Finally, to discretize and encode the selected centrality measure, we employ the equal-width binning technique \cite{catlett1991changing}. Specifically, we generate a $d$-dimensional one-hot encoded feature vector $\hat{\bf x}_{i,*}$ for node $i$ in $G_*$ such that 1 is assigned to the $\ceil{\frac{c_{i,*}}{w}}$-th element of $\hat{\bf x}_{i,*}$, where $\lceil \cdot \rceil$ denotes the ceiling operator. Here, $ w = \frac{c_{\max}-c_{\min}}{d}$, where $c_{\max}$ and $c_{\min}$ represent the highest and lowest centrality values among all nodes' centralities in the sets $\mathcal{V}_s$ and $\mathcal{V}_t$, respectively. The adjustment of the binning width $w$ determines the degree of stringency applied to enforce consistency in feature vectors of the cross-network node pair, thereby yielding a {\it trade-off between consistency and expressiveness}. That is, a higher value of $w$ leads to a lower value of $d$, representing high consistency and low expressiveness accordingly. This trade-off is empirically investigated in Appendix H.1.


\subsubsection{GNN-Based Embedding Similarity Calculation}
We turn to describing how to calculate the multi-layer embedding similarity matrix after learning node representations via GNNs along with the augmented node feature matrix $\hat{\mathcal{X}}_*$. 
Our empirical finding indicates that na\"ively concatenating augmented node features $\hat{\mathcal{X}}_*$ and original node features $\mathcal{X}_*$ reveals unsatisfactory alignment performance. As a sophisticated alternative, we train two \textit{separate} GNN models, $GNN_{\theta_1}$ and $GNN_{\theta_2}$, using a layer-wise reconstruction loss \cite{park2023power} to learn more distinguishable node representations in \eqref{hid_rep_withorigin} and \eqref{hid_rep} :
\begin{equation}
\begin{aligned}
   \label{layer-wise reconstruction loss}
    &\mathcal{L}_1 = \sum_{* \in \{s,t\}}\sum_l \left\|\tilde{\mathbf{D}}_*^{(l)-\frac{1}{2}}\tilde{\mathbf{A}}_*^{(l)}\tilde{\mathbf{D}}_*^{(l)-\frac{1}{2}}-\mathbf{H}_*^{(l)}{\mathbf{H}_*^{(l)~\top}}\right\|_F \\
    &\mathcal{L}_2 = \sum_{* \in \{s,t\}}\sum_l \left\|\tilde{\mathbf{D}}_*^{(l)-\frac{1}{2}}\tilde{\mathbf{A}}_*^{(l)}\tilde{\mathbf{D}}_*^{(l)-\frac{1}{2}}-\hat{\mathbf{H}}_*^{(l)}{\hat{\mathbf{H}}_*^{(l)~\top}}\right\|_F,
\end{aligned}
\end{equation}
where $\tilde{\mathbf{A}}_*^{(l)} =\sum_{k=1}^l\bar{\mathbf{A}}_*^{k}$ where $\bar{\mathbf{A}}_*=\mathbf{A}_*+\mathbf{I}_*$ is the adjacency matrix with self-connections in which $\mathbf{I}_*$ is the identity matrix; $\tilde{\mathbf{D}}_*^{(l)}$ is a diagonal matrix whose $(i,i)$-th element is $\tilde{\mathbf{D}}_*^{(l)}(i,i) = \sum_j\tilde{\mathbf{A}}_*^{(l)}(i,j)$, where $\tilde{\mathbf{A}}_*^{(l)}(i,j)$ is the $(i,j)$-th element of $\tilde{\mathbf{A}}_*^{(l)}$; and $\|\cdot\|_F$ is the Frobenius norm of a matrix. 

Finally, we compute the multi-layer embedding similarity matrix  ${\bf S}_{emb}$ in \eqref{emb_sim}. Here, the hyperparameter $\lambda$ plays a crucial role in determining the alignment accuracy. The optimal value of $\lambda$ can be selected based on the consistency of the original node features in two given networks. We empirically show the impact of $\lambda$ in Appendix H.1.

\subsubsection{Gradual NA with ACN Similarity Calculation}
\begin{figure}
    \centering
    \includegraphics[scale=0.5]{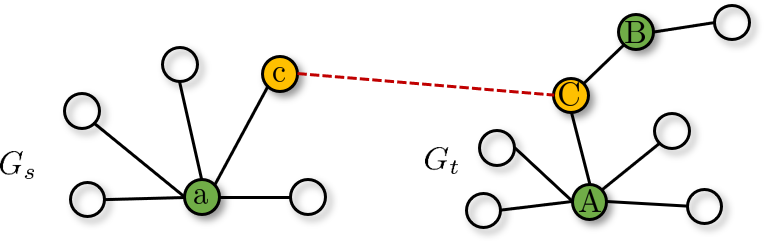}
    \caption{An instance that describes two networks $G_s$ and $G_t$ to be aligned. Here, the red dashed line denotes a matched node pair (c,C) in the first iteration of gradual NA.}
    \label{fig:example1}
\end{figure}

Next, we explain how to gradually discover node pairs by leveraging the information of ACNs and the multi-layer embedding similarity. We remark that prior methods adopted the Jaccard index \cite{du2019joint} and the Tversky similarity \cite{park2023power} to calculate the similarities between cross-network nodes. Nonetheless, the methods mentioned above tend to assign high similarity scores to node pairs with low degrees, which is not desirable in real-world applications. 
\begin{example}
We show the case where using the Jaccard index between cross-network nodes $(u,v)$ for NA does not precisely capture the  structural consistency of $(u,v)$. Let ${\bf S}_{Jac}^{(i)} (u,v)$ denote the Jaccard index between pair $(u,v)$ in the $i$th-iteration, which is expressed as follows:
\begin{equation}
\begin{aligned}
\label{Jaccard sim}
&\mathbf{S}_{Jac}^{(i)}(u,v) = \frac{ACN_{u,v}^{(i)}}{|\pi^{(i)}(\mathcal{N}_u) \cup \mathcal{N}_v|}, 
\end{aligned}
\end{equation}
where $\pi^{(i)}(\mathcal{N}_u)$ is the set of nodes whose elements are mapped via $\pi^{(i)}$. As illustrated in Fig. \ref{fig:example1}, we consider two networks $G_s$ and $G_t$ with $n_s=6$ and $n_t=8$. When there is a single matched pair $(c,C)$ in the first iteration, we have ${\bf S}_J^{(1)}(a,A)=\frac{1}{9}$ and ${\bf S}_{Jac}^{(1)}(a,B)=\frac{1}{6}$. This indicates that, although the structural consistency of $(a,A)$ is higher than that of $(a,B)$, using the Jaccard index would favor aligning node $a$ with node $B$ rather than $A$ since a node pair having a higher degree would result in a smaller similarity score.

\end{example}

To overcome this problem, we present our new measure, {\it ACN similarity} in \eqref{ACN_sim}, which focuses on the number of ACNs itself without normalization to the total number of neighbors unlike the Jaccard index in \eqref{Jaccard sim}. 

Finally, using the similarity matrix ${\bf S}^{(i)}$ in \eqref{overall_sim}, our \textsf{Grad-Align+} gradually discovers node pairs while making full use of the information from our CNFA.

\subsection{Theoretical Analysis}
In this subsection, we aim to thoroughly provide the theoretical analyses for validating the effectiveness of core components in \textsf{Grad-Align+}. Additionally, we refer to Appendix F for the complexity analysis of \textsf{Grad-Align+}. It is worth noting that our conference version \cite{park2022gradalign+} has never shown such theoretical validations of our method.
\subsubsection{Expressiveness Analysis of Augmented Node Features}
To verify the enhanced {\it expressiveness} of node representations via GNNs using CNFA, we use the Weisfeiler--Lehman (WL) graph isomorphism test (WL test) \cite{weisfeiler1968reduction}. This test recursively aggregates color labels of nodes and their neighborhoods, and hashes the aggregated labels into unique new color labels, which has a close connection with the message passing mechanism in GNNs \cite{gilmer2017neural,DBLP:conf/nips/HamiltonYL17,xu2018powerful}.

To establish a theoretical foundation that connects GNNs and the WL test, we describe the hidden representation of node $u$ at the $l$-th GNN layer, denoted by ${\bf h}_u^{(l)}$, as follows:
\begin{equation}
\label{GNNs_1}
    \mathbf{h}_{u,*}^{(l)} = \phi(\mathbf{h}_{u,*}^{(l-1)}, g(\{\textbf{h}_{i,*}^{(l-1)}:i \in \mathcal{N}(u)\})),
\end{equation}
where $\phi$ and $g$ are update and aggregation functions, respectively, which are assumed to be injective.
On the other hand, the WL test applies a predetermined injective hash function $\xi$ to update the node labels $r_{u,*}^{(l)}$:
\begin{equation}
\label{WL_eq}
    r_{u,*}^{(l)} = \xi(r_{u,*}^{(l-1)}, \{r_{i,*}^{(l-1)}:i \in \mathcal{N}(u)\}).
\end{equation}
Using \eqref{GNNs_1} and \eqref{WL_eq}, we are ready to show that, at the $l$-th GNN layer, the expressiveness of the hidden representations $\mathbf{\hat{h}}_{u,*}^{(l)}$ using augmented node features is always higher than that of the hidden representations $\mathbf{h}_{u,*}^{(l)}$ without any CNFA. The proof is provided in Appendix E.
\begin{theorem} 
\label{theorem_expressiveness}
Suppose that $\phi$, $g$, and $f_w$ are all injective. We also assume that a mapping function between a node degree and a color label in the WL test is injective. Then, when degree centrality is adopted in CNFA, in the $l$-th GNN layer, the expressive power of $\mathbf{\hat{h}}_{u,*}^{(l)}$ is higher than that of $\mathbf{h}_{u,*}^{(l)}$ for all nodes $u\in G_*$.
\end{theorem}
We refer to Appendix E for the proof. Theorem \ref{theorem_expressiveness} states that incorporating the augmented node features into the GNN model enhances the expressiveness of a node's representation. It is worthwhile to claim that, for some datasets, the actual expressive power of node representations facilitated by other centrality measures is even higher than that of the degree centrality used for the sake of analytical tractability. We empirically validate this claim in Section \ref{rq1}.

\subsubsection{Influence of CNFA on Vector Representations}

We analyze how augmented feature vectors influence the resulting vector representations via GNNs. For a ground truth node pair $(u,v)$, it is highly probable to have a small $\| \hat{\bf h}_{u,s}-\hat{\bf h}_{u,t} \|_2$ in the embedding space, where $\| \cdot\|_2$ is the $L_2$-norm of a vector (or a matrix). For ease of analysis, given ground truth node pairs $(u,v)$, we make the following two assumptions: 1) $\mathbb{E}[\|\hat{\bf x}_{u,s}-\hat{\bf x}_{v,t}\|_2]$ is negligibly small, that is, augmented features are consistent, and 2) networks $G_*$ are $r$-regular as in \cite{ma2022meta}. Now, we delve into our theoretical finding by establishing the following theorem. 

\begin{theorem}
\label{theorem 4.2} 
Consider the pre-activation output of the 1-layer GCN model in which the weight matrix ${\bf W}$ is shared. Suppose that $\mathbb{E}[\|\hat{\bf x}_{u,s}-\hat{\bf x}_{v,t}\|_2] \le \epsilon$ for an arbitrarily small $\epsilon>0$ by the node feature consistency assumption given a ground truth node pair $(u,v)$ for $u\in\mathcal{V}_s$ and $v\in\mathcal{V}_t$. Then, it follows that 
\begin{equation}
\begin{aligned}
    &\mathbb{E}[\|\hat{\bf h}_{u,s}-\hat{\bf h}_{v,t}\|_2] \\
    &\le\epsilon + \frac{1}{r}\|\mathbf{W}\|_2(\mathbb{E}[|\mathcal{N}_u-\mathcal{S}_{uv}|] + \mathbb{E}[|\mathcal{N}_v-\mathcal{S}_{uv}|]),
\end{aligned}
\label{eq15}
\end{equation}
where $\mathcal{N}_u$ is the set of neighbors of node $u$ and $\mathcal{S}_{uv}$ denotes the set of ACNs between the node pair $(u,v)$.
\end{theorem}

We refer to Appendix E for the proof. Following Theorem \ref{theorem 4.2}, when $|\mathcal{N}_s| = |\mathcal{N}_v| = |\mathcal{S}_{uv}|$ (i.e., the best case), $\mathbb{E}[\|\hat{\bf h}_{u,s}-\hat{\bf h}_{v,t}\|_2]$ is bounded by $\epsilon$. This indicates that, for a ground truth node pair $(u,v)$, if all the neighbors of $(u,v)$ are ACNs,\footnote{This is sensical since ground truth node pairs tend to share lots of ACNs in real-world networks \cite{chen2020cone}.} then, $\mathbb{E}[\|\hat{\bf h}_{u,s}-\hat{\bf h}_{v,t}\|_2]$ is arbitrarily small.

\section{Experimental Evaluation}\label{section 5}
In this section, we provide our extensive empirical study to answer the following five key research questions (RQs). 

\begin{itemize}
    \item \textit{RQ1.} How does the choice of centrality in CNFA affect the performance of \textsf{Grad-Align+}?
    \item \textit{RQ2.} How much does the \textsf{Grad-Align+} method improve performance over state-of-the-art NA methods?    
    \item \textit{RQ3.} How much does CNFA influence the performance of various NA methods?
    \item \textit{RQ4.} How do key parameters affect the performance of \textsf{Grad-Align+}?
    \item \textit{RQ5.} How robust is our \textsf{Grad-Align+} method to the structural/feature noise of given networks?

\end{itemize}

Additionally, we provide experimental results related to scalability and sensitivity analysis in Appendix H.
\begin{table}[t!]
\footnotesize
\begin{center}
\begin{tabular}{cccccc}
\hline
\multicolumn{2}{c}{Datasets} & {\begin{tabular}[c]{@{}c@{}} $|\mathcal{V}_*|$ \end{tabular}} & {\begin{tabular}[c]{@{}c@{}} $|\mathcal{E}_*|$\end{tabular}} & {\begin{tabular}[c]{@{}c@{}} $d_o$\end{tabular}} & {\begin{tabular}[c]{@{}c@{}} $M$\end{tabular}}          \\ \hline
\multirow{2}{*}{\begin{tabular}[c]{@{}c@{}}Facebook \\Twitter\end{tabular}}    & $G_s$  & 1,043 & 4,734 & - & \multirow{2}{*}{1,043}\\
                                                                        & $G_t$ & 1,043 & 4,860  & - &                       \\ \hline
\multirow{2}{*}{\begin{tabular}[c]{@{}c@{}}Douban Online \\ Douban Offline\end{tabular}} & $G_s$  & 3,906 & 8,164 & 538  & \multirow{2}{*}{1,118} \\
                                                                        & $G_t$ & 1,118 & 1,511 & 538  &                       \\ \hline
\multirow{2}{*}{\begin{tabular}[c]{@{}c@{}}Allmovie \\ IMDb\end{tabular}}   & $G_s$  & 6,011  & 124,709 & 14 & \multirow{2}{*}{5,176} \\
                                                                        & $G_t$ & 5,713 & 119,073 & 14  &                       \\ \hline
\multirow{2}{*}{\begin{tabular}[c]{@{}c@{}}DBLP\\(Its noisy version)\end{tabular}}    & $G_s$  & 2,151& 6,006 & 8  & \multirow{2}{*}{2,151}\\
                                                                        & $G_t$ & 2,151 & 6,007  & 8 &                       \\ \hline
\multirow{2}{*}{\begin{tabular}[c]{@{}c@{}}PPI\\(Its noisy version)\end{tabular}}                                                   & $G_s$  & 1,819  & 5,676 & 20  & \multirow{2}{*}{1,258}\\
                                                                        & $G_t$ & 1,819 & 5,672 & 20  &                       \\ \hline
\multirow{2}{*}{\begin{tabular}[c]{@{}c@{}}Foursquare\\(Its noisy version)\end{tabular}}                                                   & $G_s$  & 17,355  & 132,208 & -  & \multirow{2}{*}{17,355} \\
                                                                        & $G_t$ & 17,355 & 131,018 & -  &                       \\ \hline
                                                                    
\end{tabular}
\caption{Statistics of the six datasets used in our experiments. Here, Fb-Tw and Foursquare datasets are non-attributed networks, where the original node feature is unavailable.}
\label{datasettable}
\end{center}
\vspace{-3mm}
\end{table}

\subsection{Experimental Settings}
In this subsection, we summarize an overview of the experimental settings used in our study. Detailed descriptions of the datasets, competitors, and evaluation metrics can be found in Appendix G.

\subsubsection{Datasets} We conduct experiments on six benchmark datasets that are widely used for evaluating the performance of NA \cite{trung2020adaptive, park2023power,park2022gradalign+, man2016predict, du2019joint}, including three real-world datasets, namely Facebook vs. Twitter (Fb-Tw), Douban Online vs. Douban Offline (Do-Doff), and Allmovie vs. IMDb (Am-ID), and three synthetic datasets, namely DBLP, PPI, and Foursquare. Table \ref{datasettable} summarizes the key statistics of each dataset, including the number of nodes ($|\mathcal{V}_*|$), the number of edges ($|\mathcal{E}_*|$), the dimensionality of original node features ($d_o$), and the number of ground truth node pairs ($M$).

\subsubsection{Competitors} We compare our proposed \textsf{Grad-Align+} with seven state-of-the-art NA methods, including PALE~\cite{man2016predict}, FINAL~\cite{zhang2016final}, CENALP~\cite{du2019joint}, GAlign~\cite{trung2020adaptive}, Cone-Align~\cite{chen2020cone}, JORA~\cite{zheng2022jora}, and Grad-Align~\cite{park2023power}.

\subsubsection{Performance Metrics}
To assess the performance of \textsf{Grad-Align+} method and seven state-of-the-art NA methods, as the most popular metric, we adopt the {\em alignment accuracy} \cite{zhang2016final, du2019joint,park2023power}, denoted as {\em Acc}, which quantifies the proportion of correct node correspondences out of the total $M$ correspondences. We also adopt another performance metric, {\em Precision@q} (also known as $Success@q$) \cite{zhou2018deeplink, trung2020adaptive, zhang2016final}, which indicates whether there is the true positive matching identity in top-$q$ candidates.

\subsubsection{Implementation Details}
\label{sec 5.4}
We first describe the implementation details of \textsf{Grad-Align+}. The key parameters in \textsf{Grad-Align+} are listed as follows: The dimensionality of augmented node features ($d$); The exponent of $ACN_{u,v}^{(i)}$ in \eqref{ACN_sim} ($p$); The proportion (\%) of prior seed node pairs out of $M$ ground truth node pairs ($t$); The parameter balancing between two embedding similarity matrices in \eqref{emb_sim} ($\lambda$). Unless otherwise stated, the above parameters are set to $d=15$, $p=1.5$, $t=0$, and $\lambda=0.3$ as pivot values. We use GIN \cite{xu2018powerful}, which exhibits the best NA performance among well-known GNN models including GCN \cite{DBLP:conf/iclr/KipfW17} and GraphSAGE \cite{DBLP:conf/nips/HamiltonYL17} (see \cite{park2023power} for more details). For CENALP, Grad-Align, and \textsf{Grad-Align+}, we set the number of iterations in gradual NA as 10. We note that the other hyperparameters in all seven state-of-the-art methods are tuned to provide the best performance. We basically assume {\it unsupervised settings} where prior anchor links are unavailable. Nevertheless, for the NA methods that should operate on supervision data (i.e., FINAL and JORA), we use randomly selected 5\% of prior anchor links as supervision data although our method is handicapped accordingly. For the FB-Tw dataset where the original node feature is unavailable, we use all-ones vectors $\mathbf{1} \in \mathbb{R}^{1 \times n_s}$ and $\mathbf{1} \in \mathbb{R}^{1 \times n_t}$, as in \cite{park2023power}. Unless otherwise specified, we generate a noisy version of each synthetic dataset by removing 10\% of edges and replacing 10\% of node features with zeros. We conduct each experiment over 10 different random seeds to evaluate the average performance. All experiments are carried out with Intel (R) 12-Core (TM) i7-9700K CPUs @ 3.60 GHz and 32GB RAM.








 \subsection{Experimental Results and Analyses}
 \label{sec 5e}
 For RQ1 and RQ4, we show the results on the three real-world datasets, since the results on the three synthetic datasets showed similar tendencies. For RQ3, we use the two non-attributed network datasets to analyze the impact of CNFA in networks with no original node features. For RQ5, we use the three synthetic datasets to evaluate the performance under noisy conditions. We refer to Appendix H for further experimental results that were not shown due to page limitations.

\begin{table}[t!]
\scriptsize
\centering
\setlength\tabcolsep{8pt} 
\scalebox{1}{
\begin{tabular}{cccccc}
\toprule 
Dataset & Centrality& $s_c$ & $\mathbb{E}[\| \Delta\mathbf{\hat{\mathbf{x}}} \|_2]$& $\mathbb{E}[\|\Delta\mathbf{\hat{\mathbf{h}}}\|_2]$ & \textit{Acc}\\
\midrule
\multirow{6}{*}{\rotatebox{0}{Fb-Tw}} 
  & Degree & \textbf{0.35}&\underline{0.33}  & \underline{3.57} & \textbf{91.26} \\
  & Eigenvector &0.03 & 0.45  & 5.55 & 0.02 \\
  & Katz & \underline{0.33}&\textbf{0.16} & \textbf{2.63} & \underline{89.18} \\
  & Betweenness & 0.28& 0.34  & 4.74 & 81.22 \\
  & PageRank & \underline{0.33}& 0.36  & 4.55 & 84.92 \\
  & Closeness &0.32 & 0.43  & 4.17 & 71.02 \\
\midrule
\multirow{6}{*}{\rotatebox{0}{Do-Doff}} 
  & Degree & \textbf{0.37} & 0.65  & 4.28 & \underline{50.72} \\
  & Eigenvector & 0.35 & 0.49  & 4.32 & 47.76 \\
  & Katz &0.31 & \textbf{0.44} & \textbf{4.14} & 45.97 \\
  & Betweenness & 0.35& 0.62  & \underline{4.17} & 48.66 \\
  & PageRank & \underline{0.36} &  \underline{0.46} & \underline{4.17} & \textbf{51.23} \\
  & Closeness &0.35 & 1.46  & 5.02 & 41.23 \\
\midrule
\multirow{6}{*}{\rotatebox{0}{Am-ID}} 
  & Degree & \underline{0.36}& 0.61  & 4.40 & 93.16 \\
  & Eigenvector & \underline{0.36}& \textbf{0.33}  & \underline{3.80} & \underline{93.34} \\  
  & Katz &0.35 & 0.46  & 3.95 & 93.22 \\
  & Betweenness &0.35 & 0.51  & 3.90 & 92.97 \\
  & PageRank &0.35 &0.75 & 4.43 & 93.12 \\
  & Closeness &\textbf{0.37} &\underline{0.34}  & \textbf{3.31} & \textbf{93.35} \\
\bottomrule
\end{tabular}}\\
\caption{Empirical analysis on the relationship among augmented node features, node representations, and performance on {\it Acc} according to six centrality measures on three real-world datasets, 
where $\Delta\mathbf{\hat{\mathbf{x}}}=\mathbf{\hat{\mathbf{x}}}_{u,s}-\mathbf{\hat{\mathbf{x}}}_{v,t}$ and  $\Delta\mathbf{\hat{\mathbf{h}}}=\mathbf{\hat{\mathbf{h}}}_{u,s}-\mathbf{\hat{\mathbf{h}}}_{v,t}$ given ground truth node pairs $(u,v)$ for $u\in\mathcal{V}_s$ and $v\in\mathcal{V}_t$. Here, the best and second-best performers are highlighted in {\bf bold} and \underline{underline}, respectively.}

\label{Q1table}
\end{table}

\begin{table*}[t!]
\centering
\scalebox{1.15}{
\begin{tabular}{cccccccccc}
\toprule 
Method & Metric &Fb-Tw & Do-Doff& Am-ID & DBLP  & PPI & Foursquare & \textit{Unsup.} \\
\midrule 
\multirow{3}{*}{\rotatebox{0}{PALE}} 
  & {\em Acc} & 0.5923  & 0.1052 & 0.5323 & 0.5152 &  0.5888 & 0.3122\\
  & $Precision@5$ & 0.6851  & 0.1783 & 0.6244 &  0.5833 & 0.6334 & 0.3564  \\
  & $Precision@10$ & 0.7303  & 0.2338 & 0.7519 &  0.6032  & 0.6912 & 0.3887\\
\midrule
\multirow{3}{*}{\rotatebox{0}{FINAL}} 
  & {\em Acc} & 0.6328  & 0.2773 & 0.6125  & 0.6775 & 0.8756& 0.4211& \\
  & $Precision@5$ & 0.6475  & 0.4358  &0.7592  &0.6895 & 0.9132&0.4577 &\\
  & $Precision@10$ & 0.7253  & 0.5824  & 0.8152  &0.7211 & 0.9358&0.4912 & \\
\midrule
\multirow{3}{*}{\rotatebox{0}{CENALP}} 
  & {\em Acc} & \underline{0.9105} &0.0235 & 0.4238  &0.6124  & 0.9442&  0.7624&\\
  & $Precision@5$ & \underline{0.9352}  &  0.0571 & 0.5721  &0.7135   &  0.9521& 0.7924 &\checkmark \\
  & $Precision@10$ & \underline{0.9405} & 0.1130 & 0.7154  &0.7745  &  0.9588&  0.8133&\\
\midrule
\multirow{3}{*}{\rotatebox{0}{GAlign}} 
  & {\em Acc} & 0.0513  & 0.2568 & 0.7364 &0.9126 & 0.9123 & 0.0241& \\
  & $Precision@5$ & 0.0422 & 0.5233 & 0.8101   &0.9340  & 0.9332 & 0.0287 &\checkmark \\
  & $Precision@10$ & 0.0612  & 0.6324 & 0.8749  &0.9623  & 0.9622 & 0.0397& \\
\midrule
\multirow{3}{*}{\rotatebox{0}{CONE-Align}} 
  & {\em Acc} & 0.5964 &0.0100 & 0.1012  &0.3758  &  0.6383 & \underline{0.8342}&\\
  & $Precision@5$ & 0.6779  &  0.0118 & 0.2044 & 0.5372& 0.7482 & \underline{0.8512} &\checkmark \\
  & $Precision@10$ & 0.7430 & 0.0168 & 0.2859  & 0.6852  & 0.8296 & \underline{0.8733}&\\
\midrule
\multirow{3}{*}{\rotatebox{0}{JORA}} 
  & {\em Acc} & 0.0806  & 0.2968 & 0.8314 &\underline{0.9326}  & 0.9347 & 0.0844& \\
  & $Precision@5$ & 0.0911 & 0.5633 & 0.9009   &\underline{0.9562} & 0.9499 & 0.0913 &  \\
  & $Precision@10$ & 0.1033  & \underline{0.6736} & 0.9249  &\underline{0.9713}  & 0.9715 & 0.1077& \\
\midrule
\multirow{3}{*}{\rotatebox{0}{Grad-Align}} 
  & {\em Acc} &  0.0218 & \underline{0.2987} & \underline{0.8316}  & 0.9115  & \underline{0.9650} & 0.0231 & \\
  & $Precision@5$ & 0.0325  &\underline{0.5707} & \underline{0.9101}  & 0.9475  & \underline{0.9832} & 0.0274 &\checkmark \\
  & $Precision@10$ & 0.0421 &  0.6494  & \underline{0.9308}  &0.9495  & \underline{0.9898} & 0.0313 &\\
\midrule
\multirow{3}{*}{\rotatebox{0}{\textsf{Grad-Align+}}} 
  & {\em Acc} & \textbf{0.9126}  &  \textbf{0.4866} & \textbf{0.9318}   & \textbf{0.9837} &\textbf{0.9973} & \textbf{0.9572} &\\
  & $Precision@5$ & \textbf{0.9336}  &\textbf{0.6522} &\textbf{0.9640} & \textbf{0.9995} &\textbf{0.9989} & \textbf{0.9668} &\checkmark\\
  & $Precision@10$ & \textbf{0.9462} & \textbf{0.7369} &\textbf{0.9879} & \textbf{1.0000}  & \textbf{0.9995} & \textbf{0.9772} &\\
\bottomrule
\end{tabular}}
\caption{Performance comparison among \textsf{Grad-Align+} and seven state-of-the-art NA methods in terms of the {\it Acc} and {\it Precision@q} for $q\in\{5,10\}$. Here,
\textit{Unsup.} represents the methods that are run unsupervisedly without
any prior anchor links. The best and second-best performers are highlighted in {\bf bold} and \underline{underline}, respectively.}
\label{Q3table}
\end{table*}
\subsubsection{Impact of CNFA \textbf{(RQ1)}}
\label{rq1}
We empirically show how our centrality selection score in \eqref{centrality selection score} and node representations behave according to different nodes' centrality measures used for NFA. Table \ref{Q1table} summarizes an empirical analysis on the relationship among augmented node features, node representations, and performance on NA according to six centrality measures on three real-world datasets, where $\mathbb{E}[\| \mathbf{\hat{\mathbf{x}}}_{u,s}-\mathbf{\hat{\mathbf{x}}}_{v,t} \|_2]$, $\mathbb{E}[\| \mathbf{\hat{\mathbf{h}}}_{u,s} - \mathbf{\hat{\mathbf{h}}}_{v,t} \|_2]$, and {\it Acc} are shown for each centrality given ground truth node pairs $(u,v)$ for $u\in\mathcal{V}_s$ and $v\in\mathcal{V}_t$. Our findings are as follows:
\begin{itemize}
\item The selection of a particular centrality measure does not always dominate others in terms of expressiveness of node representations and {\it Acc}. Thus, one needs to choose an appropriate centrality measure depending on the dataset.
\item Higher $s_c$ tend to result in a higher alignment accuracy. This indicates that $s_c$ indeed precisely captures the key properties of NFA (i.e., expressiveness and consistency).
\item A lower $\mathbb{E}[\| \mathbf{\hat{\mathbf{x}}}_{u,s}-\mathbf{\hat{\mathbf{x}}}_{v,t} \|_2]$ tend to result in a lower $\mathbb{E}[\| \mathbf{\hat{\mathbf{h}}}_{u,s} - \mathbf{\hat{\mathbf{h}}}_{v,t} \|_2]$, which finally lead to a higher alignment accuracy. This is consistent with Theorem \ref{theorem 4.2} in the sense that $\mathbb{E}[\|\hat{\bf h}_{u,s}-\hat{\bf h}_{u,t}\|_2]$ is bounded by small $\mathbb{E}[\|\hat{\bf x}_{u,s}-\hat{\bf x}_{u,t}\|_2]$, conditioned that most of neighbors of pair $(u,v)$ are ACNs.
\item Except the eigenvector centrality on Fb-Tw, satisfactory performance can be achieved regardless of centrality measures and datasets. The use of eigenvector centrality on Fb-Tw results in undesirable performance since the centrality tends to concentrate on only a few nodes in this particular network \cite{rodrigues2019network}, thus weakening the expressive power of node representations.
\end{itemize}
We remark that our conference version \cite{park2022gradalign+} did not present any centrality selection scheme and rather performed post-hoc evaluations based on selecting the best centrality measure in terms of maximizing the alignment accuracy using ground truth cross-network node pairs, which is {\it impractical} since ground truth node correspondences are unavailable beforehand. In contrast, in this study, CNFA is performed without knowing ground truth node correspondences.

\definecolor{linecol1}{rgb}{1.0, 0.5, 0.0}
\definecolor{linecol2}{rgb}{0.1, 0.6, 0.0}
\definecolor{linecol3}{rgb}{0.2, 0.4, 0.8}
\definecolor{linecol4}{rgb}{0.2, 0.1, 0.7}
%


\subsubsection{Comparison with State-of-the-Art NA Methods \textbf{(RQ2)}}
Table \ref{Q3table} presents the performance comparison among \textsf{Grad-Align+} and seven NA competitors with respect to the {\it Acc} and {\it Precision@q} for $q\in\{5,10\}$ using three real-world and three synthetic datasets. We would like to make the following insightful observations:
\begin{itemize}
    \item \textsf{Grad-Align+} consistently and significantly outperforms all the competitors regardless of the datasets and the performance metrics while showing gains of \textsf{Grad-Align+} ($X$) over the second-best performer ($Y$) up to 69.80\% in terms of {\it Acc}, where the improvement rate (\%) is given by $\frac{X-Y}{Y}\times 100$ (\%). Notably, despite being handicapped, \textsf{Grad-Align+} achieves such dramatic gains without utilizing any supervision data. 
    \item However, the second-best performer varies depending on the datasets, which implies that one does not dominate other competitors.
    \item GAlign, JORA, and Grad-Align, which are GNN-based competitors, perform poorly on the datasets without node features such as Fb-Tw and Foursquare. In contrast, \textsf{Grad-Align+} achieves state-of-the-art performance on those datasets with the aid of CNFA. This indicates that the augmented node features indeed play a crucial role in generating precise representations of each node, thus resulting in a substantial performance improvement.
    \item On Do-Doff having very weak structural consistency \cite{du2019joint}, existing NA methods relying heavily on structural information (e.g., CENALP and CONE-Align) tend to perform unsatisfactorily. In this dataset, \textsf{Grad-Align+} outperforms the second-best competitor by large margins up to 69.80\% in terms of the {\it Acc}.
\end{itemize}

\begin{table}[t!]
\footnotesize
\centering
\setlength\tabcolsep{8pt} 
\scalebox{1}{
\begin{tabular}{cccccc} 
\toprule 
\multicolumn{1}{c}{Dataset} & {Methods} & {\begin{tabular}[c]{@{}c@{}} w/o \\ CNFA \end{tabular}} & {\begin{tabular}[c]{@{}c@{}} w/ \\ CNFA \end{tabular}} & {\begin{tabular}[c]{@{}c@{}} Gain \\ ($\%$)\end{tabular}} \\
\midrule
\multirow{5}{*}{\rotatebox{0}{Fb-Tw}} 
  & FINAL & \underline{0.6328}  & 0.7138 &  12.80\\
  & CENALP & \textbf{0.9105}  & \textbf{0.9208} &  1.13\\
  & GAlign & 0.0513  & 0.8032 &  1465.70\\
  & JORA & 0.0806  & 0.8531 &  1058.43\\
  & Grad-Align & 0.0248  &0.8705 & \underline{3410.08} \\
  & \textsf{Grad-Align+} & 0.0257 & \underline{0.9126} &  \textbf{3450.67} \\
\midrule
\multirow{5}{*}{\rotatebox{0}{Foursquare}} 
  & FINAL & \underline{0.4211}  & 0.7345 & 74.42 \\
  & CENALP & \textbf{0.7624}  & 0.8064 & 5.77\\
  & GAlign & 0.0331  &  0.6854 & 1970.70\\
  & JORA & 0.0844  & 0.8776 &  1039.81\\
  & Grad-Align & 0.0234 &  \underline{0.8972}&  \underline{3734.19}\\
  & \textsf{Grad-Align+} & 0.0228 & \textbf{0.9572} &  \textbf{4098.24}\\
\midrule
\end{tabular}}
\caption{Performance comparison of NA methods with and without CNFA in terms of the {\it Acc} on the non-attributed network datasets. Here, the best and second-best performers are highlighted in bold and underline, respectively.}
\label{Q6table}
\end{table}


\subsubsection{Impact of CNFA on Other NA Methods \textbf{(RQ3)}}
\label{rq4}
The CNFA phase can also be integrated into other NA methods that can leverage node feature information (i.e., FINAL, CENALP, GAlign, and Grad-Align). To investigate the impact of CNFA in such NA methods, we conduct an ablation study by removing the component of CNFA. Table \ref{Q6table} summarizes the evaluation results in terms of the {\it Acc} using the non-attributed network datasets (i.e., Fb-Tw and Foursquare). Our findings are as follows. 
\begin{itemize}
    \item The integration of CNFA into existing NA methods consistently exhibits superior performance compared to its counterpart (i.e., the cases without CNFA).
    \item Especially, dramatic gains up to x40.98 over the cases without CNFA are achieved when GNN-based NA methods such as GAlign, JORA, Grad-Align, and \textsf{Grad-Align+} are employed.
    \item Since the accuracy of {\it interim} discovery of node correspondences is critical in gradually discovering node pairs, CNFA can most benefit both Grad-Align and \textsf{Grad-Align+} by helping find correct node pairs, especially in the early stage of gradual node matching. On the other hand, although CENALP also gradually discovers node pairs, the gain is quite marginal since CENALP does not make use of node feature information for network embedding \cite{du2019joint}.
\end{itemize}
\definecolor{barcol1}{rgb}{0.2, 0.8, 1.0}
\definecolor{barcol2}{rgb}{0.1, 0.6, 0.0}
\definecolor{barcol3}{rgb}{0.7, 0.2, 0.0}
\definecolor{barcol4}{rgb}{0.8, 0.8, 0.8}
\definecolor{barcol5}{rgb}{0.1, 0.1, 0.8}
\definecolor{barcol6}{rgb}{0.1, 0.8, 0.1}
\definecolor{barcol7}{rgb}{0.8, 0.1, 0.1}
\definecolor{barcol8}{rgb}{0.1, 0.1, 0.1}

\begin{figure}[t!]
    \centering
    \pgfplotsset{compat=1.18,
    /pgfplots/ybar legend/.style={
    /pgfplots/legend image code/.code={%
       \draw[##1,/tikz/.cd,yshift=-0.25em]
        (0cm,0cm) rectangle (3pt,0.8em);},
   },
}
    \begin{tikzpicture}
    \begin{axis}[
        width  = 0.89\columnwidth,
        height = 3.3cm,
        major x tick style = transparent,
        ytick ={0, 0.2, 0.4, ..., 1},
        ybar = 0,
        scaled y ticks = true,
        bar width=0.035*\columnwidth,
        ylabel = $Acc$,
        symbolic x coords={Fb-Tw, Do-Doff, Am-ID},
        xtick = data,
        enlarge x limits=0.2,
        legend cell align=center,
        legend columns = 2,
        legend style={at={(0.5,1.1)}, anchor=south,legend columns=3,font=\footnotesize}
    ]
        \addplot[style={black,fill=barcol1, mark=none}]
            coordinates {(Fb-Tw, 0.91) (Do-Doff, 0.50) (Am-ID, 0.93) };
        \addplot[style={black,fill=barcol2,mark=none}]
            coordinates {(Fb-Tw, 0.03) (Do-Doff, 0.48) (Am-ID, 0.87)};
        \addplot[style={black,fill=barcol3,mark=none}]
            coordinates {(Fb-Tw, 0.86) (Do-Doff, 0.36) (Am-ID, 0.83)};
        \addplot[style={black,fill=barcol4,mark=none}]
            coordinates {(Fb-Tw, 0.42) (Do-Doff, 0.33) (Am-ID, 0.73)};
        \legend{\textsf{Grad-Align+},\textsf{Grad-Align+-(c)}, \textsf{Grad-Align+-(a)},\textsf{Grad-Align+-(g)}}
    
    \end{axis}
\end{tikzpicture}
    \caption{Performance comparison among \textsf{Grad-Align+} and its three variants in terms of the $Acc$.}
    \label{ablation_plot}
\end{figure}
\subsubsection{Module Ablation Study \textbf{(RQ4)}}

In order to examine what role each component plays in the success of the proposed \textsf{Grad-Align+} method, we perform an ablation study by removing each component in our method.
\begin{itemize}
    \item \textsf{Grad-Align+}: This corresponds to the original \textsf{Grad-Align+} method
without removing any components.
    \item \textsf{Grad-Align+(-c)}: The component of CNFA is removed. Only original node features are utilized to compute the multi-layer embedding similarity.
    \item \textsf{Grad-Align+(-a)}: The ACN similarity used during the gradual NA is replaced with the Jaccard index \cite{du2019joint}.
    \item \textsf{Grad-Align+(-g)}: Gradual NA is deactivated. That is, all node pairs are discovered at once without iterative updates of the ACN similarity matrix.
\end{itemize}
The performance comparison among the original \textsf{Grad-Align+} and its three variants is presented in Fig. \ref{ablation_plot} with respect to {\it Acc} using three real-world datasets.  We observe the following:

\begin{itemize}
\item The original \textsf{Grad-Align+} method always exhibits potential gains over other variants, which demonstrates that each component plays a critical role together in discovering node correspondences.
\item Especially for the Fb-Tw dataset, the gain of \textsf{Grad-Align+} over \textsf{Grad-Align+(-c)} is significant. This is consistent with our previous finding that GNN-based NA methods perform poorly in non-attributed networks \cite{park2023power,trung2020adaptive}, but \textsf{Grad-Align+} shows outstanding performance due to the CNFA component.
\end{itemize}

\begin{figure}[t!]
\pgfplotsset{footnotesize,samples=10}
\centering
\begin{tikzpicture}
\begin{axis}[
legend columns=4,
legend entries={\textsf{Grad-Align+},Grad-Align, JORA, Cone-Align,GAlign,CENALP,FINAL, PALE},
legend to name=named,
xmax=20,xmin=1,ymin= 0,ymax=1,
xlabel=(a) DBLP,ylabel={\em Acc}, width = 3.2cm,  height = 3.4cm,
xmin=10,xmax=50,ymin= 0,ymax=1,
xtick={10,20,...,50},ytick={0,0.2,0.4,...,1}]
    \addplot+[color=black] coordinates{(10,0.9838)(20,0.9426)(30,0.8910)(40,0.8665)(50,0.7996)};
    \addplot+[color=orange] coordinates{(10,0.9126)(20,0.8897)(30,0.8452)(40,0.7842)(50,0.6948)};
    \addplot coordinates{(10,0.9336)(20,0.8720)(30,0.8114)(40,0.7252)(50,0.6124)};
    \addplot+[color=purple] coordinates{(10,0.3758)(20,0.3324)(30,0.2644)(40,0.1955)(50,0.1683)};
    \addplot coordinates{(10,0.9126)(20,0.7316)(30,0.6811)(40,0.6234)(50,0.5977)};
    \addplot coordinates{(10,0.6124)(20,0.5985)(30,0.4230)(40,0.2985)(50,0.2652)};
    \addplot coordinates{(10,0.6775)(20,0.5324)(30,0.4724)(40,0.3685)(50,0.3374)};
    \addplot coordinates{(10,0.4152)(20,0.3224)(30,0.2624)(40,0.2385)(50,0.1874)};

\end{axis}
\end{tikzpicture}
\begin{tikzpicture}
\begin{axis}[
xmax=20,xmin=1,ymin= 0,ymax=1,
xlabel=(b) PPI, width = 3.2cm,  height = 3.4cm,
xmin=10,xmax=50,ymin= 0.2,ymax=1,
xtick={10,20,...,50},ytick={0.2,0.4,...,1}]
    \addplot+[color=black] coordinates{(10,0.9973)(20,0.9126)(30,0.8810)(40,0.8475)(50,0.7996)};
    \addplot+[color=orange] coordinates{(10,0.9650)(20,0.8897)(30,0.8452)(40,0.7842)(50,0.6948)};
    \addplot coordinates{(10,0.9347)(20,0.8757)(30,0.7122)(40,0.6252)(50,0.5124)};
    \addplot+[color=purple] coordinates{(10,0.6383)(20,0.6121)(30,0.5844)(40,0.4655)(50,0.3983)};
    \addplot coordinates{(10,0.9123)(20,0.8716)(30,0.7911)(40,0.7634)(50,0.5377)};
    \addplot coordinates{(10,0.9442)(20,0.7985)(30,0.6230)(40,0.4985)(50,0.3652)};
    \addplot coordinates{(10,0.8756)(20,0.6324)(30,0.5724)(40,0.3685)(50,0.2374)};
    \addplot coordinates{(10,0.5118)(20,0.4524)(30,0.2924)(40,0.2185)(50,0.1974)};
\end{axis}
\end{tikzpicture}
\begin{tikzpicture}
\begin{axis}[
xmax=20,xmin=1,ymin= 0,ymax=1,
xlabel=(c) Foursquare, width = 3.2cm,  height = 3.4cm,
xmin=10,xmax=50,ymin= 0,ymax=1,
xtick={10,20,...,50},ytick={0,0.2,0.4,...,1}]
    \addplot+[color=black] coordinates{(10,0.9612)(20,0.9126)(30,0.8810)(40,0.8375)(50,0.7996)};
    \addplot+[color=orange] coordinates{(10,0.021)(20,0.020)(30,0.019)(40,0.018)(50,0.017)};
    \addplot coordinates{(10,0.0815)(20,0.0744)(30,0.0712)(40,0.0621)(50,0.0314)};
    \addplot+[color=purple] coordinates{(10,0.832)(20,0.762)(30,0.704)(40,0.617)(50,0.461)};
    \addplot coordinates{(10,0.022)(20,0.022)(30,0.020)(40,0.018)(50,0.015)};
    \addplot coordinates{(10,0.762)(20,0.734)(30,0.675)(40,0.5856)(50,0.526)};
    \addplot coordinates{(10,0.4211)(20,0.3675)(30,0.3114)(40,0.2885)(50,0.2074)};
    \addplot coordinates{(10,0.2711)(20,0.2111)(30,0.1422)(40,0.1022)(50,0.0974)};

\end{axis}
\end{tikzpicture}
\\
\ref{named}
\caption{Alignment accuracy according to different levels of the structural noise (\%) on the three synthetic datasets.}
\label{Q4plot_structure}
\end{figure}
\begin{figure}[t!]
\pgfplotsset{footnotesize,samples=10}
\centering
\begin{tikzpicture}
\begin{axis}[
legend columns=3, 
legend entries={\textsf{Grad-Align+},Grad-Align, JORA, GAlign,CENALP,FINAL},
legend to name=named,
xlabel=(a) DBLP,ylabel={\em Acc},  width = 3.5cm, height = 3.4cm,
xmin=10,xmax=50,ymin= 0.2,ymax=1,
xtick={10,20,...,50},ytick={0.2,0.4,...,1}]
    \addplot+[color=black] coordinates{(10,0.9838)(20,0.9426)(30,0.9310)(40,0.8965)(50,0.8496)};
    \addplot+[color=orange] coordinates{(10,0.9126)(20,0.8897)(30,0.8452)(40,0.7842)(50,0.6948)};
    \addplot coordinates{(10,0.9326)(20,0.8516)(30,0.8111)(40,0.7234)(50,0.6077)};
    \addplot coordinates{(10,0.9126)(20,0.7316)(30,0.6811)(40,0.6234)(50,0.5977)};
    \addplot coordinates{(10,0.6124)(20,0.5985)(30,0.4730)(40,0.3985)(50,0.3652)};
    \addplot coordinates{(10,0.6775)(20,0.5324)(30,0.4724)(40,0.3685)(50,0.3374)};
\end{axis}
\end{tikzpicture}
\begin{tikzpicture}
\begin{axis}[
xlabel=(b) PPI, width = 3.5cm, height = 3.4cm,
xmin=10,xmax=50,ymin= 0.4,ymax=1,
xtick={10,20,...,50},ytick={0.4,0.6,...,1}]
    \addplot+[color=black] coordinates{(10,0.9973)(20,0.9626)(30,0.9310)(40,0.9175)(50,0.9096)};
    \addplot+[color=orange] coordinates{(10,0.9650)(20,0.8897)(30,0.8452)(40,0.7842)(50,0.6948)};
    \addplot coordinates{(10,0.9347)(20,0.8216)(30,0.7611)(40,0.7134)(50,0.6377)};
    \addplot coordinates{(10,0.9123)(20,0.8716)(30,0.7911)(40,0.7634)(50,0.6777)};
    \addplot coordinates{(10,0.9442)(20,0.8985)(30,0.7730)(40,0.6385)(50,0.6052)};
    \addplot coordinates{(10,0.8756)(20,0.7224)(30,0.6724)(40,0.5885)(50,0.4974)};
\end{axis}
\end{tikzpicture}
\\
\ref{named}
\caption{Alignment accuracy according to different levels of the feature noise (\%) on the two synthetic datasets.}
\label{Q4plot_att}
\end{figure}
\subsubsection{Robustness to Network Noises \textbf{(RQ5)}}

We now compare our \textsf{Grad-Align+} method to the seven NA competitors in two more difficult settings that often occur in real-world networks: 1) the case in which a significant portion of edges in two given networks $G_s$ and $G_t$ are removed and 2) the case in which a large portion of node features in $G_s$ and $G_t$ are missing and replaced with zeroes. The performance is presented according to different levels of structural and feature noises in Figs. \ref{Q4plot_structure} and \ref{Q4plot_att}, respectively. Our findings are as follows.
\begin{itemize}
    \item {\bf Scenario 1.} In Fig. \ref{Q4plot_structure}, we show how {\it Acc} behaves according to the random removal of $\{10,\cdots,50\}\%$ of existing edges in each of three synthetic datasets. While the performance tends to degrade with an increasing level of the structural noise for all the methods, \textsf{Grad-Align+} consistently achieves superior performance compared to all competitors for all noise levels.
    \item {\bf Scenario 2.} In Fig. \ref{Q4plot_att}, we show how {\it Acc} behaves according to the random replacement of $\{10,\cdots,50\}\%$ of node features with zeros. Due to the fact that node features are unavailable on Foursquare, the performance is presented using two other synthetic datasets. Similarly, as in Scenario 1, all methods exhibit a performance decrease with an increasing level of the feature noise. However, \textsf{Grad-Align+} consistently outperforms all competitors by large margins while being quite robust to node feature noises. This robustness is attributed to the use of both CNFA and ACN similarity calculation in \textsf{Grad-Align+}, which leverage the structural information to enhance the robustness against node feature noises.
\end{itemize}

\section{Concluding Remarks}\label{section 6}

In this paper, we aimed to devise a new methodology that substantially improves the performance of NA in unsupervised settings where prior anchor links and node features are unavailable. Towards this goal, we proposed \textsf{Grad-Align+}, the high-quality NA method that judiciously integrates the GNN model trained along with augmented node features based on CNFA into the gradual node matching framework. Through extensive experiments on three real-world and three synthetic datasets, we demonstrated (a) the impact of CNFA on the representation learning and {\it Acc}, (b) the superiority of \textsf{Grad-Align+} over seven state-of-the-art NA methods while showing dramatic gains up to 69.80\% in terms of the {\it Acc} compared to the best NA competitor, (c) the impact of CNFA on other NA methods, (d) the importance of each component in \textsf{Grad-Align+} via an ablation study, and (e) the robustness of \textsf{Grad-Align+} to two types of network noises. Our theoretical analyses also uncovered (a) the benefits of using CNFA for enhancing the expressiveness of representations via GNNs, (b) the effect of centrality selection in CNFA on node representations, and (c) the scalability of \textsf{Grad-Align+}.

Potential avenues of future research include the design of more sophisticated yet effective NFA for NA while sufficing three design principles such as expressiveness, permutation invariance, and consistency. It is expected that this will enable more accurate discovery of node correspondences in the early stages of gradual NA, thus potentially boosting the NA performance.

\ifCLASSOPTIONcompsoc
  \section*{Acknowledgments}
\else
  \section*{Acknowledgment}
\fi

This work was supported by the National Research Foundation of Korea (NRF) grant funded by the Korea government (MSIT) (No. 2021R1A2C3004345, No. RS-2023-00220762). The material in this paper was presented in part at the ACM International Conference on Information and Knowledge Management, Atlanta, GA, October 2022 \cite{park2022gradalign+}.

\ifCLASSOPTIONcaptionsoff
  \newpage
\fi




\bibliographystyle{IEEEtran}
\bibliography{IEEEabrv,1.Citation_list}

\newpage

\ifCLASSINFOpdf
\else
\fi
\hyphenation{op-tical net-works semi-conduc-tor IEEE-Xplore}

%
\title{Centrality-Based Node Feature Augmentation for Robust Network Alignment (Appendix)}

\markboth{Submitted to IEEE Transactions on Network Science and Engineering}%
{Shell \MakeLowercase{\textit{et al.}}: Bare Demo of IEEEtran.cls for Computer Society Journals}
%



\maketitle

\IEEEdisplaynontitleabstractindextext

%
\IEEEpeerreviewmaketitle

\appendices
\section{Notations}
Table \ref{NotationTable} summarizes the notation that is used in the main manuscript as well as the following sections in this material. This notation is formally defined in the main manuscript when we introduce our methodology and the technical details.
\begin{table}[t]
\scriptsize
    \centering
    \begin{tabular*}{0.99\columnwidth}{ll}
        \toprule
        \textbf{Notation} & \textbf{Description} \\
        \midrule
        \rule{0pt}{7pt}$G_s$ & Source network\\
        \rule{0pt}{7pt}$G_t$ & Target network\\
        \rule{0pt}{7pt}$\mathcal{V}_*$ & Set of nodes in $G_*$\\
        \rule{0pt}{7pt}$\mathcal{E}_*$ & Set of edges in $G_*$\\
        \rule{0pt}{7pt}$\mathcal{X}_*$ &Set of feature vectors of nodes in $\mathcal{V}_*$\\
        \rule{0pt}{7pt}$\mathcal{\hat{\mathcal{X}}}_*$ &Set of augmented feature vectors of nodes in $\mathcal{V}_*$\\
        \rule{0pt}{7pt}$n_*$ & Number of nodes in $G_*$\\
        \rule{0pt}{7pt}$\pi^{(i)}$ & One-to-one node mapping at the $i$-th iteration\\
        \rule{0pt}{7pt}$M$ & Total number of ground truth node pairs\\
        \rule{0pt}{7pt}$\tilde{\mathcal{V}}_*^{(i)}$ & Set of seed nodes in $G_*$ up to the $i$-th iteration\\
        \rule{0pt}{7pt}$\hat{\mathcal{V}}_*^{(i)}$ & Set of newly aligned nodes in $G_*$ at the $i$-th iteration\\
        \rule{0pt}{7pt}$\tilde{\mathcal{V}}_*^{(0)}$ & Set of prior seed nodes in $G_*$\\
        \rule{0pt}{7pt}$\mathbf{\mathbf{H}}_*^{(l)}$ & $l$-th GNN layer's hidden representation matrix from $\mathcal{X}_*$ in $G_*$\\
        \rule{0pt}{7pt}$\mathbf{\hat{\mathbf{H}}}_*^{(l)}$ & $l$-th GNN layer's hidden representation matrix from $\mathcal{\hat{X}}_*$ in $G_*$\\
        \rule{0pt}{7pt}$\mathbf{S}_{emb}$ & Multi-layer embedding similarity matrix\\
        \rule{0pt}{7pt}$\mathbf{S}_{ACN}^{(i)}$ & ACN similarity matrix at the $i$-th iteration\\
        \bottomrule
    \end{tabular*}
    \caption{Summary of notations. Here, the subscript $*$ represents $s$ and $t$ for source and target networks, respectively. }
    \label{NotationTable}
\end{table}

\section{Expressive power of centrality measures}
\begin{figure}
    \centering
    \includegraphics[scale=0.42]{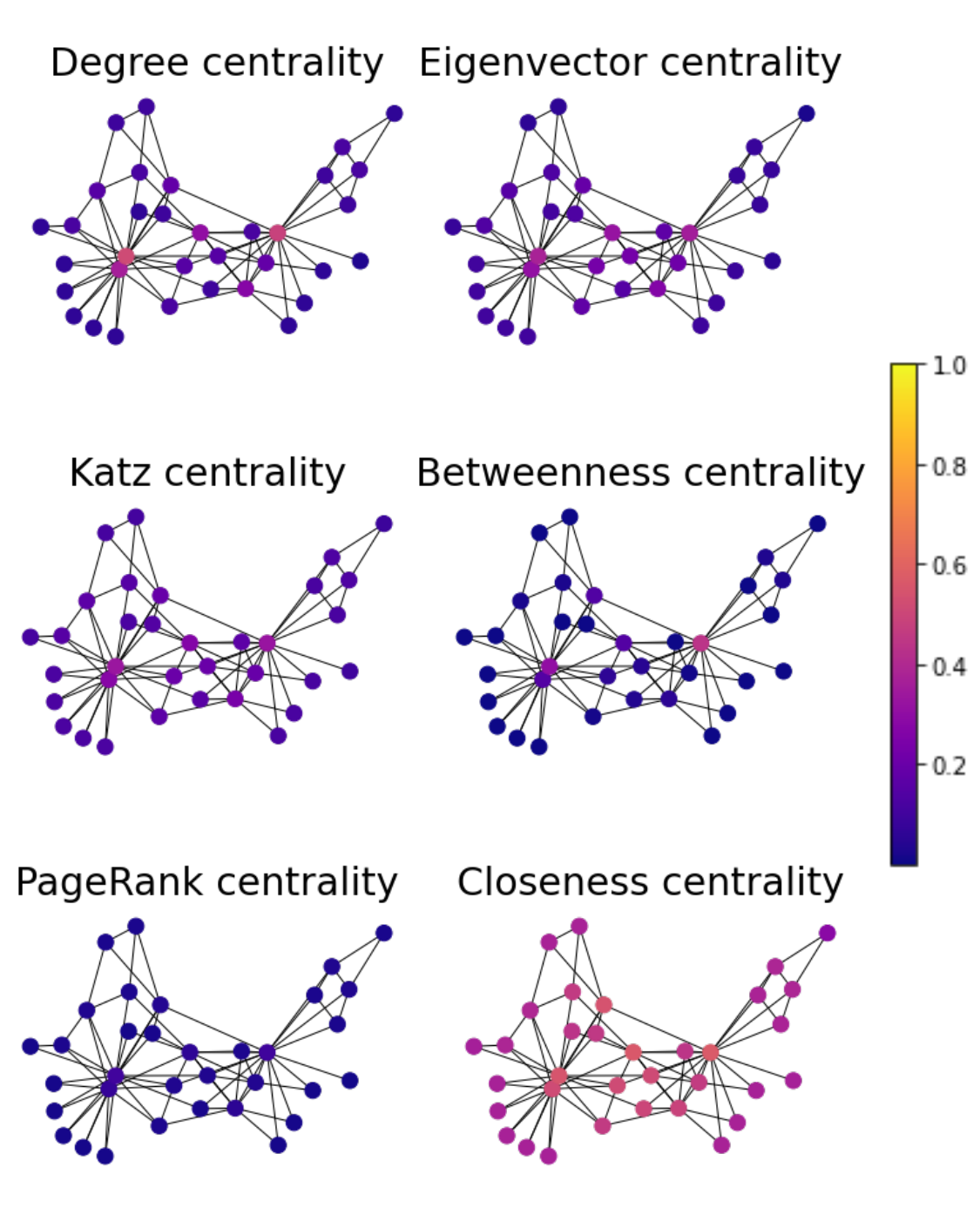}
    \caption{A visualization of different centrality measures on Zachary’s Karate club. Here, different node colors represent different centrality values.}
    \label{fig:example_karate}
\end{figure}  
We show a toy example that examines the expressive power of different centrality measures. To this end, Fig. \ref{fig:example_karate} visualizes the nodes exhibiting different values for six centrality measures in the Zachary's Karate club network \cite{zachary1977information}. From the figure, it is observed that PageRank and closeness centralities are less expressive than others in the network.

\section{Related Work}
Our proposed method in this paper is related to three research fields, namely NFA for downstream ML tasks on graphs, NA only with structural information, and NA with both feature and structural information.

{\bf NFA for downstream ML tasks on graphs.}
In general, NFA for ML tasks on graphs was mostly used for creating high-quality training data or improving the capability of model generalization \cite{shorten2019survey,zhao2022graph}. LA-GNN \cite{liu2022local} was designed to generate node features using a generative model that learns the distributions of node features of neighbors. FLAG \cite{kong2020flag} was proposed by augmenting node features for adversarial training. SR+DR \cite{song2021topological} was proposed by generating topology embeddings as input node features, which are extracted based on random walk-based representation learning. GCA \cite{zhu2021graph} utilized NFA for graph  contrastive learning. NodeAug \cite{wang2020nodeaug} was proposed by replacing node features based on the node feature distribution of two-hop neighborhoods.

{\bf NA only with structural information.} 
When only structural information of networks is available, NA methods were carried out in the sense of leveraging the {\it structural consistency}, while identifying the same nodes across different networks based on their topological characteristics. For example, IsoRank \cite{singh2008global} was designed by propagating pairwise node similarity and structural consistency to discover the node correspondence. NetAlign \cite{bayati2009algorithms} formulated the NA problem as an integer quadratic program. BIG-ALIGN \cite{koutra2013big} was designed by using the alternating projected gradient descent approach for solving a constrained optimization problem to find permutation matrices. Recently, network-embedding-aided NA has become in the spotlight due to its scalability and high expressiveness. PALE \cite{man2016predict} utilized the first and the second-order proximities of nodes in the embedding space, and employed a multi-layer perceptron (MLP) to capture the nonlinear relationship between embeddings from two different networks. DeepLink \cite{zhou2018deeplink} was designed by using an unbiased random walk for the node embeddings of two given networks. CONE-Align \cite{chen2020cone} presented a way of aligning the embedding subspaces by modeling intra-network proximity with node embeddings and then using these embeddings to match nodes across networks. cM$^2$NE\cite{xiong2021contrastive} was developed for multiplex network embeddings by utilizing multiple structural views of given networks. WL-Align \cite{liu2023wl} proposed to use cross-network Weisfeiler--Lehman relabeling. CCNE \cite{zhang2024collaborative} was developed by collaboratively learning intra- and inter-network structures in the same latent space. However, we may pose a practical challenge on the aforementioned methods since, if structural consistency is violated, then additional information is required for accurate node matching.

{\bf NA with both feature and structural information.}
Node features have been shown to be useful in discovering node correspondences \cite{park2023power, zhang2016final}. For instance, user profile attributes (e.g., name, affiliation, and description) can aid in aligning the same user across multiple social networks \cite{zhang2018mego2vec}. Thus, incorporating feature information into NA methods can complement the solution to overcome the violation of the structural consistency assumption  \cite{zhang2016final}. Numerous methods \cite{zhang2016final,heimann2018regal,du2019joint,trung2020adaptive,park2023power,gao2021unsupervised,zheng2022jora,huynh2021network} have been developed to perform the NA task along with node feature information. REGAL \cite{heimann2018regal} presented a low-rank implicit approximation of a similarity matrix that takes into account both structural similarity and feature agreement between nodes in two given networks. FINAL \cite{zhang2016final} leveraged not only node feature information but also edge feature information. CENALP \cite{du2019joint} showed how to incrementally discover node pairs while jointly performing the link prediction and NA tasks. As GNNs \cite{DBLP:conf/iclr/KipfW17, DBLP:conf/nips/HamiltonYL17,xu2018powerful,velivckovic2017graph} have emerged as a powerful means to extract useful low-dimensional features in attributed networks with node features, GAlign \cite{trung2020adaptive} was developed by exploiting the multi-order nature of graph convolutional network (GCN) \cite{DBLP:conf/iclr/KipfW17} for solving NA. WAlign \cite{gao2021unsupervised} was designed by minimizing the Wasserstein distance between embeddings of nodes via lightweight GCN. JORA \cite{zheng2022jora} was developed by jointly optimizing the representation learning and alignment learning components using GNNs as a backbone model. HCT \cite{sun2023towards} was developed by explicitly utilizing high-order structural consistency, which is formulated based on edge orbits for unsupervised NA. CPUM \cite{tang2023identifying} presented an adaptive graph attention network for jointly modeling node features and topology in social networks. Grad-Align \cite{park2023power} showed how to gradually discover node pairs by calculating both the multi-layer embedding similarity of nodes using GNNs and the similarity between cross-network nodes using the information of the so-called {\it ACN}.

{\bf Discussion.} 
 Despite the recent active research on NFA and NA, existing studies face several practical challenges. First, conventional NFA techniques for ML tasks on graphs were mostly designed to create additional training data to improve the capability of model generalization \cite{shorten2019survey,zhao2022graph} whose purpose is not appropriate for improving the performance of NA. For example, the aforementioned NFA methods are based on perturbation or sampling \cite{kong2020flag, zhu2021graph, wang2020nodeaug}, and are \textit{not permutation-invariant} \cite{song2021topological}, which can lead to a significant deterioration in NA performance. Furthermore, while some existing NA methods \cite{zhang2016final,tang2023identifying} in part use NFA for NA, they restrict their usage to merely employing degree centrality, resulting in suboptimal performance. In other words, NFA for enhancing NA performance remains largely underexplored. Second, despite the above-described contributions, the satisfactory performance of existing NA methods relies on additional information, such as 1) prior anchor links \cite{man2016predict,zhou2018deeplink,du2019joint, zhang2016final} and 2) node features \cite{trung2020adaptive,zhang2016final, park2023power, heimann2018regal}, which are not always available in real-world applications. In this case, properly augmented node features for NA will relieve the problem of such a high dependency on the additional information in existing NA methods, which motivates us to design a new NA method that makes full use of NFA. To the best of our knowledge, such an attempt has not been conducted in the literature as long as NA is concerned.

\section{Pseudocode of \textsf{Grad-Align+}}
The overall procedure of the proposed \textsf{Grad-Align+} method along with our CNFA module is summarized in Algorithm \ref{mainalgorithm} and Algorithm \ref{CNFAalgorithm}, respectively, where $\tilde{\mathcal{V}}_s^{(i)}$ and $\tilde{\mathcal{V}}_t^{(i)}$ denote the set of aligned nodes (including supervision data if any) in $G_s$ and $G_t$, respectively, up to the $i$-th iteration; $M$ is the total number of ground truth cross-network node pairs to be discovered; and $N$ is the number of node pairs to be discovered for each gradual step.

   \begin{algorithm}[t]
\caption{: \textsf{Grad-Align+}}
\label{mainalgorithm}
 \begin{algorithmic}[1]
  \renewcommand{\algorithmicrequire}{\textbf{Input:}}
  \renewcommand{\algorithmicensure}{\textbf{Output:}}
  \REQUIRE $G_s=(\mathbf{A}_s,\mathcal{X}_s)$, $G_t=(\mathbf{A}_t,\mathcal{X}_t)$, $\tilde{\mathcal{V}}_s^{(0)}$,$\tilde{\mathcal{V}}_t^{(0)}$
  \ENSURE $\pi^{(\ceil{\frac{M}{N}}+1)}$
  \STATE \textbf{Initialization: } $\theta \leftarrow \text{random initialization}$; $i \leftarrow 1$
  \STATE /* CNFA */
  \STATE  $\mathcal{\hat{X}}_s, \mathcal{\hat{X}}_t\leftarrow \text{CNFA}(\mathbf{A_s}, \mathbf{A_t})$
  \STATE /* GNN-based embedding similarity calculation */
  
  \WHILE {not converged}
  \STATE  $\mathbf{H}^{(1)}_s, \mathbf{H}^{(2)}_s, ..., \mathbf{H}^{(L)}_s\leftarrow \textsf{GNN}_{\theta_1}(\mathbf{A_s}, \mathcal{X}_s)$
  \STATE  $\mathbf{H}^{(1)}_t, \mathbf{H}^{(2)}_t, ..., \mathbf{H}^{(L)}_t\leftarrow \textsf{GNN}_{\theta_1}(\mathbf{A_t},\mathcal{X}_t)$
  \STATE $\mathcal{L} \leftarrow$ \text{layer-wise reconstruction loss} $\mathcal{L}_1$
  \STATE Update $\theta_1$ by taking one step of gradient descent
  \ENDWHILE
  
  \WHILE {not converged}
  \STATE  $\mathbf{\hat{H}}^{(1)}_s, \mathbf{\hat{H}}^{(2)}_s, ..., \mathbf{\hat{H}}^{(L)}_s\leftarrow \textsf{GNN}_{\theta_2}(\mathbf{A_s}, \mathcal{\hat{X}}_s)$
  \STATE  $\mathbf{\hat{H}}^{(1)}_t, \mathbf{\hat{H}}^{(2)}_t, ..., \mathbf{\hat{H}}^{(L)}_t\leftarrow \textsf{GNN}_{\theta_2}(\mathbf{A_t},\mathcal{\hat{X}}_t)$
  \STATE $\mathcal{L} \leftarrow$ \text{layer-wise reconstruction loss} $\mathcal{L}_2$
  \STATE Update $\theta_2$ by taking one step of gradient descent
  \ENDWHILE
  
  \STATE $\mathbf{S}_{emb} \leftarrow \sum_l \mathbf{H}_s^{(l)}{\mathbf{H}_t^{(l)~\top}} + \lambda \sum_l  \hat{\mathbf{H}}_s^{(l)}\hat{\mathbf{H}}_t^{(l)~\top}$
  \STATE /* Gradual NA with ACN similarity calculation */
  \IF {$\tilde{\mathcal{V}}_s^{(0)}=\tilde{\mathcal{V}}_t^{(0)}=\emptyset$}
    \STATE $\mathbf{S}^{(i)}\leftarrow \mathbf{S}_{emb}$
  \ELSE 
    \STATE Calculate $\mathbf{S}_{ACN}^{(i)}$
    \STATE $\mathbf{S}^{(i)}\leftarrow \mathbf{S}_{emb}\odot \mathbf{S}_{ACN}^{(i)}$
  \ENDIF
  
  \WHILE {$i\le \ceil{\frac{M}{N}}$} 
  \STATE Find $\hat{\mathcal{V}}_s^{(i)}$ and $\hat{\mathcal{V}}_t^{(i)}$ based on $\mathbf{S}^{(i)}$
  \STATE $i \leftarrow i + 1$
  \STATE Update mapping $\pi^{(i)}$: \\$\{ \pi^{(i)}(u)=v \mid u \in \hat{\mathcal{V}}_s^{(i-1)}, v \in \hat{\mathcal{V}}_t^{(i-1)}\}$
  \STATE Update ${\bf S}_{ACN}^{(i)}$
  \STATE ${\bf S}^{(i)} \leftarrow {\bf S}_{emb} \odot {\bf S}_{ACN}^{(i)}$
  \ENDWHILE
  \RETURN $\pi^{(\ceil{\frac{M}{N}}+1)}$
 \end{algorithmic}
\end{algorithm}

\begin{algorithm}[t]
\caption{CNFA}
\label{CNFAalgorithm}
 \begin{algorithmic}[1]
  \renewcommand{\algorithmicrequire}{\textbf{Input:}}
  \renewcommand{\algorithmicensure}{\textbf{Output:}}
  \REQUIRE $G_s=(\mathcal{V}_s,\mathcal{E}_s,\mathcal{X}_s)$, $G_t=(\mathcal{V}_t,\mathcal{E}_t,\mathcal{X}_t)$
  \ENSURE $\hat{\mathcal{X}}_s, \hat{\mathcal{X}}_t$
  \STATE /* Centrality selection */
    \FOR{each centrality measure}
        \STATE Calculate $c_{i,*}$ for $i \in \mathcal{V}_*$ 
        \STATE Calculate $s_c$ using (7) in the main manuscript
    \ENDFOR
    \STATE Select the centrality measure leading to the highest $s_c$
\STATE /* NFA */
  \FOR{$u \in \mathcal{V}_s$}
  \STATE $\hat{\bf x}_{u,s} \leftarrow f_w(c_{u,s})$ 
  \STATE $\mathcal{\hat{X}}_s = \mathcal{\hat{X}}_s \cup \{\hat{\bf x}_{u,s}\}$
  \ENDFOR
  \FOR{$v \in \mathcal{V}_t$}
  \STATE $\hat{\bf x}_{v,t} \leftarrow f_w(c_{v,t})$ 
  \STATE $\mathcal{\hat{X}}_t = \mathcal{\hat{X}}_t \cup \{\hat{\bf x}_{v,t}\} $
  \ENDFOR
  \RETURN $\mathcal{\hat{X}}_s,\mathcal{\hat{X}}_t$
 \end{algorithmic}
\end{algorithm}

\section{Proof of Theorems}

\setcounter{theorem}{0} 
\renewcommand{\thetheorem}{3.\arabic{theorem}} 

\begin{theorem} 
\label{theorem_expressiveness}
Suppose that $\phi$, $g$, and $f_w$ are all injective. We also assume that a mapping function between a node degree and a color label in the WL test is injective. Then, when degree centrality is adopted in CNFA, in the $l$-th GNN layer, the expressive power of $\mathbf{\hat{h}}_{u,*}^{(l)}$ is higher than that of $\mathbf{h}_{u,*}^{(l)}$ for all nodes $u\in G_*$.
\end{theorem}
\begin{proof}
We prove this theorem by the mathematical induction for all layers $l=0, 1, \cdots, L$, where $l=0$ (i.e., initial node features) and $l\ge1$ correspond to the base step and the consecutive inductive steps, respectively.
\\{\bf Base step:} There always exists an injective mapping $\varphi$ between $\mathbf{h}_{u,*}^{(l)}$ and $r_{u,*}^{(l)}$ \cite{xu2018powerful}. Thus, it follows that $\mathbf{h}^{(0)}_{u,*} =\varphi(r_{u,*}^{(0)})$. On the other hand, we have $\hat{\mathbf{h}}_{u,*}^{(0)} = \hat{\mathbf{x}}_{u,*}=f_w(c_{u,*})=f_w \circ \eta^{-1}(r_{u,*}^{(1)})=\varphi(r_{u,*}^{(1)})$, where $c_{u,*}$ is the centrality of node $u$ in $G_*$, $\eta$ is the mapping from a node degree to a color label of the node, and $\circ$ is the operator of function composition. Thus, $\hat{\mathbf{h}}_{u,*}^{(0)}=\varphi(r_{u,*}^{(1)})$ has higher expressiveness than that of $\mathbf{h}^{(0)}_{u,*}$, as it yields a one-round higher WL test outcome.
\\{\bf Inductive step:} Suppose that
\begin{align}
\hat{\bf h}_{u,*}^{(l-1)} = \varphi (r_{u,*}^{(l)}),
\end{align}
which implies that the hidden representation $\hat{\bf h}_{u,*}^{(l-1)}$ using the augmented node features is identical to a one-round higher outcome $\varphi (r_{u,*}^{(l)})$ from the WL test. Then, we have
\begin{equation}
\label{GNNs_2}
    \mathbf{\hat{h}}_{u,*}^{(l)} = \phi(\mathbf{\hat{h}}_{u,*}^{(l-1)}, g(\{\mathbf{\hat{h}}_{i,*}^{(l-1)}:i \in \mathcal{N}(u)\})),
\end{equation}
which can be transformed to
\begin{equation}
\label{GNNs_3}
    \mathbf{\hat{h}}_{u,*}^{(l)} = \phi(\varphi{(r_{u,*}^{(l)})}, g(\varphi{(r_{i,*}^{(l)})}:i \in \mathcal{N}(u)\})).
\end{equation}
Since the composition of injective functions is injective, there exists an injective function $\zeta$ such that
\begin{equation}
\label{GNNs_4}
\begin{aligned}
    \mathbf{\hat{h}}_{u,*}^{(l)} &= \zeta(r_{u,*}^{(l)}, \{r_{i,*}^{(l)}:i \in \mathcal{N}(i)\}) \\
    &= \zeta \circ \xi^{-1}\circ\xi(r_{u,*}^{(l)}, \{r_{i,*}^{(l)}:i \in \mathcal{N}(i)\}) \\
    &= \zeta \circ \xi^{-1}(r_{u,*}^{(l+1)}),
\end{aligned}
\end{equation}
where the last equality holds due to (11) in the main manuscript. As $\zeta \circ \xi^{-1}$ is also injective, we finally have $\mathbf{\hat{h}}_u^{(l)} = \varphi(r_{u,*}^{(l+1)})$, which has a one-round higher WL test outcome compared to $\mathbf{\textbf{h}}_u^{(l)} = \varphi(r_{u,*}^{(l)})$.
This completes the proof of the theorem.

\end{proof}

\begin{theorem}
\label{theorem 4.2} 
Consider the pre-activation output of the 1-layer GCN model in which the weight matrix ${\bf W}$ is shared. Suppose that $\mathbb{E}[\|\hat{\bf x}_{u,s}-\hat{\bf x}_{v,t}\|_2] \le \epsilon$ for an arbitrarily small $\epsilon>0$ by the node feature consistency assumption given a ground truth node pair $(u,v)$ for $u\in\mathcal{V}_s$ and $v\in\mathcal{V}_t$. Then, it follows that 
\begin{equation}
\begin{aligned}
    &\mathbb{E}[\|\hat{\bf h}_{u,s}-\hat{\bf h}_{v,t}\|_2] \\
    &\le\epsilon + \frac{1}{r}\|\mathbf{W}\|_2(\mathbb{E}[|\mathcal{N}_u-\mathcal{S}_{uv}|] + \mathbb{E}[|\mathcal{N}_v-\mathcal{S}_{uv}|]),
\end{aligned}
\label{eq15}
\end{equation}
where $\mathcal{N}_u$ is the set of neighbors of node $u$ and $\mathcal{S}_{uv}$ denotes the set of ACNs between the node pair $(u,v)$.
\end{theorem}

\begin{proof}
Vector representations of nodes $u$ and $v$ from GCN are expressed as follows:
\begin{equation}
\begin{aligned}
\label{eq1 in thm3.1}
&\hat{\mathbf{h}}_{u,s} = \mathbf{W}\left(\mathbf{\hat{\mathbf{x}}}_{u,s} + \frac{1}{r} \sum_{i \in \mathcal{N}_u}{\mathbf{\hat{\mathbf{x}}}_{i,s}}\right)\\
&\hat{\mathbf{h}}_{v,t} = \mathbf{W}\left(\mathbf{\hat{\mathbf{x}}}_{v,t} + \frac{1}{r}\sum_{j \in \mathcal{N}_v}{\mathbf{\hat{\mathbf{x}}}_{j,t}}\right),
\end{aligned}
\end{equation}
where $r$ is the normalizing constant in GCN. Then, we have

\begin{equation}
\begin{aligned}
&\mathbb{E}[\| \hat{\mathbf{h}}_{u,s} - \hat{\mathbf{h}}_{v,t} \|_2] \\
&= \mathbb{E}\left[\left\|\mathbf{W}\left(\mathbf{\hat{\mathbf{x}}}_{u,s} - \mathbf{\hat{\mathbf{x}}}_{v,t} + \frac{1}{r}\left(\sum_{i \in \mathcal{N}_u}{\mathbf{\hat{\mathbf{x}}}_{i,s}}- \sum_{j \in \mathcal{N}_v}{\mathbf{\hat{\mathbf{x}}}_{j,t}}\right)\right)\right\|_2\right]
\\&\le\left\|\mathbf{W}\right\|_2 \mathbb{E} \left[\left\| \mathbf{\hat{\mathbf{x}}}_{u,s} - \mathbf{\hat{\mathbf{x}}}_{v,t} + \frac{1}{r}(\sum_{i \in \mathcal{N}_u}{\mathbf{\hat{\mathbf{x}}}_{i,s}}- \sum_{j \in \mathcal{N}_v}{\mathbf{\hat{\mathbf{x}}}_{j,t}})\right\|_2\right]
\\&\le \|\mathbf{W}\|_2 \left(\epsilon + \mathbb{E} \left[\left\| \frac{1}{r}(\sum_{i \in \mathcal{N}_u}{\mathbf{\hat{\mathbf{x}}}_{i,s}}- \sum_{j \in \mathcal{N}_v}{\mathbf{\hat{\mathbf{x}}}_{j,t}})\right\|_2\right]\right)
\label{thm2_eq17}
\end{aligned}
\end{equation}
which can be simplified to \eqref{eq15}, where the first and second inequalities stem from the Cauchy--Schwarz inequality and the triangle inequality, respectively. This completes the proof of the theorem.
\end{proof}

\section{Complexity Analysis}
We theoretically analyze the computational complexity
of \textsf{Grad-Align+}. First, the computational complexity of the CNFA phase is determined depending on the choice of centrality, which can be calculated in linear time using various approximation algorithms. For example, given $G_*=(\mathcal{V}_*,\mathcal{E}_*)$, while the computational complexity of betweenness centrality is $\mathcal{O}(|\mathcal{V}_*||\mathcal{E}_*|)$ \cite{wandelt2020complex}, it can be reduced to $\mathcal{O}(|\mathcal{E}_*|)$ \cite{borassi2019kadabra}. Next, the complexity of message passing in GNNs is $\mathcal{O}(\max\{|\mathcal{E}_s|,|\mathcal{E}_t|\})$ \cite{wu2020comprehensive}. While the element-wise calculation of the similarity matrix ${\bf S}_{emb}$ in (\ref{emb_sim}) is repeated $|\mathcal{V}_s||\mathcal{V}_t|$ times \cite{du2019joint}, this process can be regarded as a constant time when parallelization is applied \cite{park2023power}. Finally, the ACN similarity calculation for a node pair $(u,v)$, corresponding to the number of ACNs, takes $|\mathcal{N}_u||\mathcal{N}_v|$, which is upper-bounded by the squared value of the maximum node degree and thus is regarded as a constant. Since the computational cost of ${\bf S}_{ACN}$ basically follows that of ${\bf S}_{emb}$, the computational complexity of \textsf{Grad-Align+} is bounded by $\mathcal{O}(\max\{|\mathcal{E}_s|,|\mathcal{E}_t|\})$. Therefore, the computational complexity of \textsf{Grad-Align+} is given by $\mathcal{O}(\max\{|\mathcal{E}_s|,|\mathcal{E}_t|\})$, which means that the computational complexity of \textsf{Grad-Align+} scales at most {\it linearly} with the maximum number of edges over two networks. We empirically validate the computational efficiency of \textsf{Grad-Align+} in Appendix G.2.

\section{Details of the Experimental Settings}

\subsection{Real-World Datasets}
\label{sec 5.1.1}
We use three real-world datasets, each of which consists of source and target networks.

\textbf{Facebook vs. Twitter} {\bf (Fb-Tw)}. The Fb-Tw dataset consists of two real-world social networks that were collected and published by \cite{cao2016bass}. User accounts are represented as nodes, and friendships between accounts are represented as edges.

\textbf{Douban Online vs. Douban Offline} {\bf (Do-Doff)}. The Douban dataset is a Chinese social network that was collected and published by \cite{zhong2012comsoc}. User accounts are represented as nodes, and friendships between accounts are represented as edges.

\textbf{Allmovie vs. IMDb} {\bf (Am-ID)}. We use two movie networks in our study: the Allmovie network,\footnote{https://www.kaggle.com/ayushkalla1/rotten-tomatoes-movie-database.} constructed from the Rotten Tomatoes (an review-aggregation website), and the IMDb network,\footnote{https://www.kaggle.com/jyoti1706/IMDBmoviesdataset.} constructed from the IMDb (an online database of movie, TV, and celebrity information). In both networks, films are represented as nodes, and two films have an edge connecting them if they share at least one actor.

\subsection{Synthetic Datasets}
\label{sec 5.1.2}
In addition to the three aforementioned real-world datasets, we synthesize network data, including one large-scale dataset with over 34k nodes, to comprehensively evaluate the performance under noisy conditions on the network structure and node features, following the approach in \cite{trung2020adaptive,du2019joint, park2023power}. Specifically, we generate a noisy version of the original network by randomly removing a certain number of edges and replacing a portion of node features with zeros, while preserving the number of ground truth cross-network node pairs. We use three synthetic datasets, including one non-attributed network and two attributed networks.

\textbf{DBLP}. The DBLP dataset, collected and published by \cite{prado2012mining}, is a co-authorship network in which authors and their academic interactions are represented as nodes and edges, respectively. Each node is associated with a feature vector indicating the number of publications in computer science conferences \cite{du2019joint}.


\textbf{PPI}. The PPI dataset \cite{breitkreutz2007biogrid} is one of widely used biological networks for NA \cite{ma2020review, saraph2014magna}. The nodes and edges represent the proteins and the interaction between them, respectively.

\textbf{Foursquare}. The Foursquare dataset, originally collected by \cite{zhang2015integrated}, is a location-based social network. The nodes and edges represent the users and the follower/followee relationships between them, respectively.

\subsection{State-of-the-Art Methods}
In this subsection, we present seven state-of-the-art NA methods for comparison.

{\bf PALE}~\cite{man2016predict}. This is a supervised NA method that performs network embedding with awareness of prior seed nodes and learns a cross-network mapping via an MLP architecture for NA.

\textbf{FINAL}~\cite{zhang2016final}. This method aligns attributed networks with node features based on the consistency principle. Specifically, FINAL utilizes three consistency conditions including topology consistency, node feature consistency, and edge feature consistency.

\textbf{CENALP}~\cite{du2019joint}. This method jointly performs NA and link prediction to increase the alignment accuracy. DeepWalk \cite{perozzi2014deepwalk} is employed to perform cross-network embeddings.

\textbf{GAlign}~\cite{trung2020adaptive}. This is an unsupervised NA method based on a multi-order GCN model using local and global structural information of networks.

\textbf{Cone-Align}~\cite{chen2020cone}. This method learns intra-network proximity with node embeddings and utilizes them to match nodes across networks via embedding subspace alignment.

\textbf{JORA}~\cite{zheng2022jora}. This is a supervised NA method that employs GNN-based network embeddings with the aid of prior anchor links and learns a cross-network mapping via an MLP architecture for NA.

\textbf{Grad-Align}~\cite{park2023power}. This method gradually discovers node pairs based on the multi-layer embedding similarity via GNNs and the Tversky similarity.

\subsection{Performance Metrics}
To assess the performance of \textsf{Grad-Align+} method and seven state-of-the-art NA methods, as the most popular metric, we adopt the {\em alignment accuracy} \cite{zhang2016final, du2019joint,park2023power}, denoted as {\em Acc}, which quantifies the proportion of correct node correspondences out of the total $M$ correspondences. We also adopt another performance metric, {\em Precision@q} (also known as $Success@q$) \cite{zhou2018deeplink, trung2020adaptive, zhang2016final}, which indicates whether there is the true positive matching identity in top-$q$ candidates and is expressed as
\begin{equation}
    Precision@q = \frac{\sum_{v_s^*\in\mathcal{V}_s} \mathds{1}_{v_t^* \in \mathcal{S}^q(v_s^*)}}{M},
\end{equation}
where $(v_s^*, v_t^*)$ is each node pair in the ground truth; $\mathcal{S}^q(v_s^*)$ indicates the set of indices of top-$q$ elements in the $v_s^*$-th row of the similarity matrix ${\bf S}^{(\ceil*{\frac{M}{N}}+1)}$; and $\mathds{1}_{\mathcal{S}_{v_s^*}^q(v_t^*)}$ is the indicator function. For node $v_s^*$, if the similarity ${\bf S}^{(\ceil*{\frac{M}{N}}+1)}(v_s^*, v_t^*)$ is ranked within the $q$-th highest values in the row ${\bf S}^{(\ceil*{\frac{M}{N}}+1)}(v_s^*,:)$ of the similarity matrix ${\bf S}^{(\ceil*{\frac{M}{N}}+1)}$, then the alignment output for $v_s^*$ is recorded as a successful case. Note that the higher the value of each of the two metrics, the better the performance.

\section{Additional Experiments}

\definecolor{linecol1}{rgb}{1.0, 0.5, 0.0}
\definecolor{linecol2}{rgb}{0.1, 0.6, 0.0}
\definecolor{linecol3}{rgb}{0.2, 0.4, 0.8}
\definecolor{linecol4}{rgb}{0.2, 0.1, 0.7}

\subsection{Effect of Key Parameters}
\label{rq2}
\begin{figure}[t!]
\pgfplotsset{footnotesize,samples=10}
\centering
\begin{tikzpicture}
\begin{axis}[
legend columns=-1,
legend entries={Fb-Tw, Do-Doff, Am-ID},
legend to name=named,
xmax=25,xmin=5,ymin= 0.2,ymax=1,
xlabel=(a) Effect of $d$.,ylabel=$Acc$, width = 4cm, height = 3.5cm,
xtick={5,10,15,20,25},ytick={0.2, 0.4, 0.6, 0.8, 1}]
    \addplot+[color=linecol1] coordinates{(5,0.805) (10,0.788) (15,0.915) (20,0.875) (25,0.825) };
    \addplot+[color=linecol2] coordinates{(5,0.49) (10,0.50) (15,0.47) (20,0.43) (25,0.39) };
    \addplot+[color=linecol3] coordinates{(5,0.885) (10,0.912) (15,0.932) (20,0.902) (25,0.875) };
\end{axis}
\end{tikzpicture}
\begin{tikzpicture}
\begin{axis}[
xmax=2.5,xmin=0.5,ymin= 0.2,ymax=1,
xlabel=(b) Effect of $p$., width = 4cm, height = 3.5cm,
xtick={0.5,1,1.5,2,2.5},ytick={0.2, 0.4, 0.6, 0.8, 1}]
    \addplot+[color=linecol1] coordinates{(0.5,0.83) (1,0.87) (1.5,0.90) (2,0.92) (2.5,0.90) };
    \addplot+[color=linecol2] coordinates{(0.5,0.47) (1,0.45) (1.5,0.50) (2,0.47) (2.5,0.43) };
    \addplot+[color=linecol3] coordinates{(0.5,0.82) (1,0.90) (1.5,0.91) (2,0.94) (2.5,0.93) };
\end{axis}
\end{tikzpicture}
\begin{tikzpicture}
\begin{axis}[
xmax=20,xmin=0,ymin= 0.2,ymax=1,
xlabel=(c) Effect of $t$.,ylabel=$Acc$, width = 4cm, height = 3.5cm,
xtick={0, 5,10,15,20},ytick={0.2, 0.4, 0.6, 0.8, 1}]
    \addplot+[color=linecol1] coordinates{(0,0.91) (5,0.94) (10,0.96) (15,0.96) (20,0.98)};
    \addplot+[color=linecol2] coordinates{(0,0.43) (5,0.49) (10,0.61) (15,0.65) (20,0.72)};
    \addplot+[color=linecol3] coordinates{(0,0.93) (5,0.96) (10,0.97) (15,0.98) (20,0.99)};
\end{axis}
\end{tikzpicture}
\begin{tikzpicture}
\begin{axis}[
xmax=0.8,xmin=0, ymin= 0, ymax=1,
xlabel=(d) Effect of $\lambda$., width = 4cm, height = 3.5cm,
xtick={0,0.2,0.4,0.6,0.8},ytick={0, 0.25, 0.5, 0.75, 1}]
    \addplot+[color=linecol1] coordinates{(0,0.053) (0.2,0.92) (0.4,0.92) (0.6,0.92) (0.8,0.92)};
    \addplot+[color=linecol2] coordinates{(0,0.48) (0.2,0.51) (0.4,0.48) (0.6,0.43) (0.8,0.39)};
    \addplot+[color=linecol3] coordinates{(0,0.86) (0.2,0.87) (0.4,0.93) (0.6,0.91) (0.8,0.81)};
\end{axis}
\end{tikzpicture}
\ref{named}
\caption{The effect of the key parameters $d,p$, and $t$ on the $Acc$ of \textsf{Grad-Align+}.}
\label{hyper_plot}
\end{figure}

We investigate the impact of parameters used in \textsf{Grad-Align+} and experimental settings, including $d$, $p$, $t$, and $\lambda$, on the NA performance. Fig. \ref{hyper_plot} shows the effect of the key parameters on {\it Acc} for the Fb-Tw, Do-Doff, and Am-ID datasets. When a hyperparameter varies so that its effect is clearly revealed, other parameters are set to the pivot values in Section 5.4 in the main manuscript. Our empirical findings are outlined below.
\begin{itemize}
    \item \textbf{The effect of $d$:} The parameter $d$ determines the dimension of augmented node feature vectors in CNFA. From Fig. \ref{hyper_plot}a, higher dimensionality does not always guarantee a higher alignment accuracy on the three datasets. For the Do-Doff dataset, having weak structural consistency \cite{du2019joint,park2023power}, the highest {\it Acc} is achieved at $d=10$, which is lower than that on Fb-Tw and Am-ID. This indicates that it is recommended to use a high value of $d$ for networks with strong structural consistency, while it is preferable to use a low value of $d$ for networks revealing structural inconsistency in order to avoid the excessive use of structural information in CNFA. This in turn demonstrates that there exists a trade-off between consistency and expressiveness.
    \item \textbf{The effect of $p$:} As shown in Fig. \ref{hyper_plot}b, the performance tends to monotonically increase with $p$ on Fb-Tw and Am-ID having strong structural consistency. In contrast, increasing $p$ on Do-Doff turns out to rather degrade the performance due to the overexploitation of ACNs during the node matching in the network exhibiting very weak structural consistency.
    \item \textbf{The effect of $t$:} From Fig. \ref{hyper_plot}c, it is obvious that the {\it Acc} monotonically increases with more supervision data. One can see that \textsf{Grad-Align+} still guarantees quite reasonable performance even in unsupervised settings (i.e., $t=0$) unlike existing NA methods.
    \item \textbf{The effect of $\lambda$:} The parameter $\lambda$ balances between the two embedding similarity matrices. As shown in Fig. \ref{hyper_plot}d, the performance increases dramatically with $\lambda>0$ on Fb-Tw, a non-attributed network dataset. In contrast, on Do-Doff and Am-ID, increasing $\lambda$ does not always show superior performance. For the Do-Doff dataset whose networks have a weak structural consistency, the highest {\it Acc} is achieved at $\lambda=0.2$. On the other hand, the Am-Id dataset achieves the highest {\it Acc} at $\lambda=0.4$. Thus, it is preferable to use a low value of $\lambda$ for networks revealing structural inconsistency.
\end{itemize}

\subsection{Validation of Scalability}

\definecolor{barcol1}{rgb}{0.2, 0.8, 1.0}
\definecolor{barcol2}{rgb}{0.1, 0.6, 0.0}
\definecolor{barcol3}{rgb}{0.7, 0.2, 0.0}
\definecolor{barcol4}{rgb}{0.8, 0.8, 0.8}
\definecolor{barcol5}{rgb}{0.1, 0.1, 0.8}
\definecolor{barcol6}{rgb}{0.1, 0.8, 0.1}
\definecolor{barcol7}{rgb}{0.8, 0.1, 0.1}
\definecolor{barcol8}{rgb}{0.1, 0.1, 0.1}

\begin{figure}[]
    \centering
    \pgfplotsset{compat=1.11,
    /pgfplots/ybar legend/.style={
    /pgfplots/legend image code/.code={%
       \draw[##1,/tikz/.cd,yshift=-0.25em]
        (0cm,0cm) rectangle (3pt,0.8em);},
   },
}
    \begin{tikzpicture}
    \begin{semilogyaxis}[
        width  = 0.95\columnwidth,
        height = 4cm,
        major x tick style = transparent,
        ymode=log,
        ybar=0,
        bar width=0.025*\columnwidth,
        ymajorgrids = true,
        ylabel = Execution time (s),
        xlabel = (a) Three real-world datasets,
        symbolic x coords={Fb-Tw, Do-Doff, Am-ID},
        xtick = data,
        enlarge x limits=0.3,
        ymin=0,
        legend cell align=center,
        legend style={at={(0.5,1.1)}, anchor=south,legend columns=4,font=\footnotesize}
    ]
    \addplot[style={black,fill=barcol1, mark=none}]
        coordinates {(Fb-Tw, 23) (Do-Doff, 132) (Am-ID, 2742)};
    \addplot[style={black,fill=barcol2, mark=none}]
        coordinates {(Fb-Tw, 21) (Do-Doff, 108) (Am-ID, 2647)};
    \addplot[style={black,fill=barcol3, mark=none}]
        coordinates {(Fb-Tw, 151) (Do-Doff, 518) (Am-ID, 1535)}; 
    \addplot[style={black,fill=barcol4, mark=none}]
        coordinates {(Fb-Tw, 128) (Do-Doff, 312) (Am-ID, 3788)};
    \addplot[style={black,fill=barcol5, mark=none}]
        coordinates {(Fb-Tw, 24) (Do-Doff, 103) (Am-ID, 455)};
    \addplot[style={black,fill=barcol6, mark=none}]
        coordinates {(Fb-Tw, 4512) (Do-Doff,10157) (Am-ID, 61801)};
    \addplot[style={black,fill=barcol7, mark=none}]
        coordinates {(Fb-Tw, 131) (Do-Doff, 135) (Am-ID, 311)};
    \addplot[style={black,fill=barcol8, mark=none}]
        coordinates {(Fb-Tw, 171) (Do-Doff, 545) (Am-ID, 2114)};

        \legend{\textsf{Grad-Align+},Grad-Align, JORA, CONE-Align, GAlign, CENALP, FINAL, PALE}
    
    \end{semilogyaxis}
\end{tikzpicture}
    \begin{tikzpicture}
    \begin{semilogyaxis}[
        width  = 0.95\columnwidth,
        height = 4cm,
        major x tick style = transparent,
        ymode=log,
        ybar=0,
        bar width=0.025*\columnwidth,
        ymajorgrids = true,
        ylabel = Execution time (s),
        xlabel = (b) Three synthetic datasets,
        symbolic x coords={DBLP, PPI, Foursquare},
        xtick = data,
        enlarge x limits=0.3,
        ymin=0,
        legend cell align=center,
        legend style={at={(0.5,1.1)}, anchor=south,legend columns=4,font=\footnotesize}
    ]
        \addplot[style={black,fill=barcol1, mark=none}]
            coordinates {(DBLP, 23) (PPI, 61) (Foursquare, 2087)};
        \addplot[style={black,fill=barcol2, mark=none}]
            coordinates {(DBLP, 23) (PPI, 58) (Foursquare, 2007)};
        \addplot[style={black,fill=barcol3, mark=none}]
            coordinates {(DBLP, 345)  (PPI, 321) (Foursquare, 7854)};
        \addplot[style={black,fill=barcol4, mark=none}]
            coordinates {(DBLP, 129) (PPI, 113) (Foursquare, 2217)};
        \addplot[style={black,fill=barcol5, mark=none}]
            coordinates {(DBLP, 47) (PPI, 102) (Foursquare, 1027)};
        \addplot[style={black,fill=barcol6, mark=none}]
            coordinates {(DBLP, 13454) (PPI, 10721) (Foursquare, 184125)};
        \addplot[style={black,fill=barcol7, mark=none}]
            coordinates {(DBLP, 301)(PPI, 177) (Foursquare, 2717)};
        \addplot[style={black,fill=barcol8, mark=none}]
            coordinates {(DBLP, 181)(PPI, 223) (Foursquare, 7112)};

        \legend{\textsf{Grad-Align+},Grad-Align, JORA, CONE-Align, GAlign, CENALP, FINAL, PALE}
    
    \end{semilogyaxis}
\end{tikzpicture}
    \caption{The runtime complexity of \textsf{Grad-Align+} and seven state-of-the-art methods on the six benchmark datasets.}
    \label{Q6plot}
\end{figure}
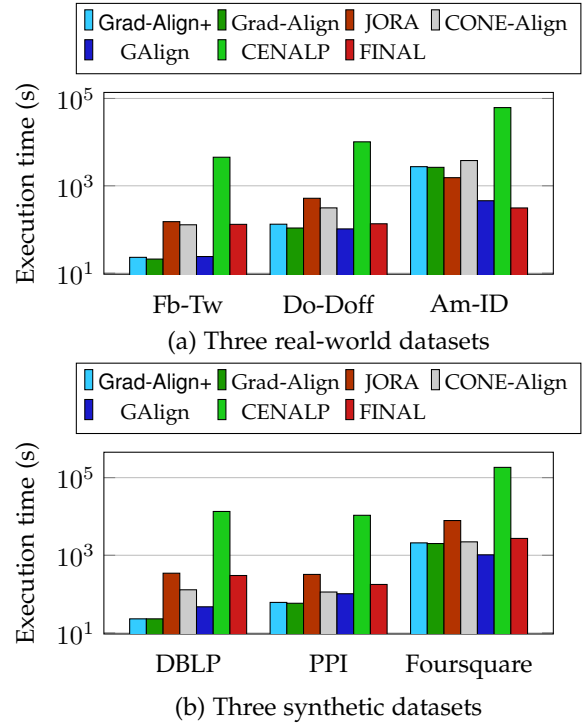

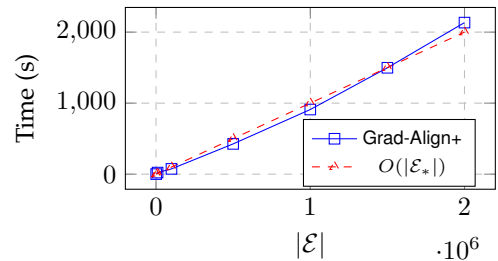
\begin{figure}[t]
\centering
\begin{tikzpicture}
    \begin{axis}[        xlabel=$|\mathcal{E}|$,        ylabel= Time (s), grid=major, grid style={dashed},  width = 6.5cm, height = 4cm , legend style={at={(0.48,0.2)},anchor=west, font=\fontsize{7}{7}\selectfont}]
        \addplot [color=blue, mark=square, legend = \textsf{CPA-LGC}]
            coordinates {
                (1000,1) (10000,22) (100000,74) (500000, 426) (1000000, 910) (1500000, 1498) (2000000, 2134) 
            };
        \addplot [ dashed, color=red,            mark=triangle, legend = $O(|\mathcal{E}|)$ ]
            coordinates {
                (1000,1) (10000,10) (100000,100) (500000, 500) (1000000, 1000) (1500000, 1500) (2000000, 2000) 
            };
    \addlegendentry{\textsf{Grad-Align+}}
    \addlegendentry{$O(|\mathcal{E}_*|)$}
    \end{axis}    
\end{tikzpicture}
\caption{Computational efficiency of \textsf{Grad-Align+}. Here, red dashed line indicates a linear scaling in $|\mathcal{E}_*|$.}
\label{scalabilityplot}
\end{figure}

First, we empirically show the average runtime complexity. We conduct experiments using the three real-world datasets (Fb-Tw, Do-Doff, and Am-ID) as well as the three synthetic datasets (DBLP, Econ, PPI, and Foursquare). Figs. \ref{Q6plot}(a) and \ref{Q6plot}(b) illustrate the execution time (in seconds) of \textsf{Grad-Align+} and seven state-of-the-art NA methods on the three real-world datasets and the three synthetic datasets, respectively. CENALP \cite{du2019joint}, which jointly performs the link prediction and NA, shows the highest runtime for all the datasets. On the other hand, the computational complexity of \textsf{Grad-Align+} is competitive to that of the seven benchmark methods. 

Second, we also empirically show the scalability of our \textsf{Grad-Align+}. We empirically validate that the computational complexity of \textsf{Grad-Align+} scales linearly with the maximum number of edges over two networks. To this end, we conduct experiments using the Erd\H{o}s--R\'enyi (ER) graph model \cite{bollobas1998random} since such generative models of graphs are able to flexibly alter the network size in order to generate networks of various sizes. From the ER graph model, we synthetically generate seven sets of two identical networks whose numbers of nodes and edges are $(|\mathcal{V}_s|,|\mathcal{E}_s|) = (|\mathcal{V}_t|,|\mathcal{E}_t|) =$\{($10^2$, $10^3$), ($5\times10^2$, $10^4$), ($10^3$, $10^5$), ($2\times10^3$, $5\times10^5$), ($4\times10^3$, $10^6$), ($6\times10^3$, $2\times10^6$)\}. The dashed line indicates a linear scaling in $|\mathcal{E}_*|$. It is seen that our empirical evaluation coincides with the theoretical analysis. Fig. \ref{scalabilityplot} shows the measured runtime of \textsf{Grad-Align+} in seconds with respect to different numbers of edges. An asymptotic dashed line is also shown in Fig. \ref{scalabilityplot}, exhibiting a trend that is consistent with our experimental result. From Fig. \ref{scalabilityplot}, we can clearly see that the computational complexity scales as $\mathcal{O}(\max\{|\mathcal{E}_*|)$.

\ifCLASSOPTIONcaptionsoff
  \newpage
\fi

\end{document}